\documentclass[aps,pra,preprint,showpacs,showkeys]{revtex4-1}

\usepackage{graphicx}
\usepackage{epsfig}
\usepackage{color}
\usepackage{setstack}
\usepackage{amsfonts,amssymb}
\usepackage{mathbbol}
\newcommand{\eqref}[1]{(\ref{#1})}
\newcommand{\eref}[1]{(\ref{#1})}
\newcommand{\sgn}{{\rm sgn}\,}

\newcommand{\re}{{\rm Re}\,}

\newtheorem{theorem}{Theorem}

\begin{document}

\title{Effective interaction in an unbalanced Fermion mixture}

\author{Christian Recher$^{1,2}$ and Heinerich Kohler$^2$}

\date{\today}

\affiliation{$^1$ \   Fakult\"at f\"ur Physik, Universit\"at Duisburg-Essen, Lotharstrasse 1,
        47048 Duisburg,     Germany  \\
        $^2$ Instituto de Ciencias Materiales de Madrid, CSIC,
           C/   Sor Juana In{\'e}s de la Cruz 3, 28049 Madrid,  Spain }
\email{christian.recher@uni-due.de}

\begin{abstract}

A one--dimensional Fermi mixture with delta--interaction is investigated in the 
limit of extreme imbalance.  In particular we consider
the cases of only one or  two minority Fermions which interact with the Fermi-sea of the majority Fermions.  We calculate dispersion relation and polaron mass for the minority Fermions as well as equal time density-density correlators. Within a cluster expansion we derive an expression for the effective interaction potential between minority Fermions.  For our calculations we use a reformulation of the exact wave functions, originally obtained by Yang and Gaudin by a nested Bethe ansatz, in terms of determinants.
\end{abstract}


\maketitle
%Para -1 ------------------------------------------------------------------------

\section{Introduction}\label{sec1}

%The study of one-dimensional many body systems has a long history
%dating back until the founding days of quantum-mechanics
%\cite{bet31}. Nowadays they serve as a paradigm in such different
%fields as........ From a theoretical point of view one-dimensional
%models are exceptional since for some interaction potentials they
%allow for an exact solution. An important class are systems which
%can be be solved by means of Bethes-Ansatz. A popular representant
%is the one-dimensional Bose-gas with point wise interaction.

%A bunch of exotic properties such as .... are related to the
%dynamics of one-dimensional models.

Imbalanced mixtures of two different species of Fermions or of Fermions and Bosons  have raised 
considerable interest during the recent years \cite{blo08,che10}. Experimentally they have 
become 
feasible in traps of laser cooled Fermi gases, where a smaller number of $^{40}$K atoms is 
moving in a sea of $^6$Li atoms or in partially polarized ensembles of  $^6$Li atoms. 
Long--standing questions of solid state physics about the coexistence of a normal and a superfluid 
in a partially polarized fermionic systems or about the emergence of a spatially varying order 
parameter giving rise to a  Fulde--Ferrell--Larkin--Ovchinnikov phase \cite{ful64,lar64} could 
thereby be addressed 
experimentally \cite{zwi06}. Since a magnetic field is expelled from charged Fermionic superfluids 
due to the Meissner effect similar experiments to date could not be performed in solid state 
systems.   

A related question arises in the highly imbalanced limit. It was predicted
\cite{com07,pro08} that in this regime a screening of the minority Fermions by a cloud of the 
majority  Fermions leads to Fermi--polaron physics. A  polaron quasiparticle peak emerging from a 
broad background in the radio--frequency adsorption spectrum  was indeed measured in 
experiments \cite{schi09,koh10}. Similar results were obtained as well in experiments with 
imbalanced  ultracold Boson mixtures \cite{spe12}.  The polaron problem in highly imbalanced 
Fermi mixtures 
was addressed theoretically by a variety of methods like Bethe--Goldstone equations \cite{che06}
diagrammatic Monte Carlo \cite{pro08},  variation of trial wave functions  \cite{com07}, functional 
renormalization group \cite{schmi11} and others \cite{baa12} both 
for weak and strong coupling.

 Recently experiments were performed on imbalanced Fermi--mixtures in one dimensional \cite
{lio10} renewing the interest in 1d systems. Since in one--dimension essentially any interaction is 
strong, the regime of strong interaction is particularly easy accessible in one--dimensional traps 
where the confinement into two directions \cite{goer01,kin04} is much stronger than in the third 
one. In 
the limit of vanishing Fermi wave vector 
$k_{\rm F}$ of the majority Fermions and diverging scattering length $c^{-1}$ with the 
dimensionless interaction strength $a=$$c / k_{\rm F} $ finite, 
the details of the interaction--potential become irrelevant and the interaction can be assumed 
delta--like. The system of spin $1/2$ Fermions with delta--interaction, often called Yang--Gaudin 
model, in one dimension is exactly solvable by Bethe's ansatz \cite{yan67, gau66}. However the 
resulting wave--functions are considerably more complicated than that of its Bosonic counterpart, 
the widely used Lieb--Liniger model \cite{lie63}.
Even in the hardcore--limit $a$ $\to \infty$, which corresponds to the Tonks--Girardeau gas in the 
Bosonic case, exact results for the dynamical density--density correlation and for the single--
particle Green's function were achieved only rather recently \cite{che04b}. 

The study of an isolated minority Fermion ( assumed a spin--up Fermion in the following) in a sea 
of spin--down Fermions was initiated  by McGuire \cite{mcg65,mcg66} and has in particular in the 
recent years attracted  much interest. The polaron problem was addressed  in \cite{gua12}. In  \cite
{brou13} numerically exact results were obtained for few particles. 

In the present work we investigate the Yang--Gaudin model  in the highly imbalanced limit. Our 
approach relies on a reformulation of the  exact many--body wave function in terms of 
determinants. This form seems at least in the imbalanced limit more suitable than the original one, 
obtained by Yang with a nested Bethe Ansatz. Using this wave function we treat  the case of two 
minority Fermions analytically. We achieve exact results for the two--point density function and 
even for higher order density correlators.  

The system's energy can be expressed exactly as a function of the free momenta of the minority 
Fermions. This yields the polaron's 
dispersion relation, from which its effective mass is derived. These results together with the 
results obtained in previous works \cite{mcg65,mcg66,rech11} 
yield a quite complete picture of the single repulsive polaron dynamics in the Yang--Gaudin 
model.   Polaron--polaron interaction is investigated within a cluster expansion of the energy. For weak coupling an effective two--body potential between the minority Fermions is derived, which is 
solely induced by the sea of majority Fermions. 
 
The paper is organized as follows. In Section \ref{sec2} we define the model and fix notation. The 
reformulation of the wave--function is described in Sec.~\ref{sec2.1}. 
The spectrum of the minority Fermions is analyzed in section \ref{sec2.3}. 
From this spectrum the effective potential between minority Fermions is derived in Sec.~\ref{sec3}. 
Density--correlation functions are investigated in Sec.~\ref{sec3.2}.

\section{Model}\label{sec2}
We consider $N+M$ Fermions on a line which interact
via a repulsive $\delta$-potential. While $N$ of the
particles are supposed to be spin-down Fermions we assume the
remaining $M$ Fermions to carry spin-up polarization. The Hamiltonian reads
\begin{eqnarray}
 \hat{H}= - \sum_{n=1}^{N} \frac{\partial^2}{\partial x_n^2} -
\sum_{m=1}^{M}\frac{\partial^2}{\partial y_m^2}
+4c \sum_{n=1}^N\sum_{m=1}^M \delta(x_n -y_m) \ , \label{eq1.0}
\end{eqnarray}
where the units
are chosen such that $\hbar=1 $ and all masses are equal to 1/2.
 Furthermore $c \geq 0$ denotes
the interaction strength and the coordinates ${\bf x}=\{x_n\}_{n=1,\ldots,N}$ and
 ${\bf y}=\{y_m\}_{m=1,\ldots,M}$ refer to the positions of the spin-down and spin-up
Fermions, respectively.

The exact eigenfunctions of the Hamiltonian \eqref{eq1.0} can be
constructed by means of Bethe's-Ansatz. For the cases
$M=1$ and $M=2$ they have been found by McGuire \cite{mcg65,
mcg66} and by Flicker \& Lieb \cite{fli67}, respectively. The
generalization of these results to an arbitrary number $M$ of
spin-up particles was overcome by Gaudin \cite{gau66} and Yang
\cite{yan67} via a nested Bethe-Ansatz. The eigenfunctions of $H$ 
constructed by this method are of the form
\begin{eqnarray}
\label{eq1.0-1}
\Psi({\bf{x}},{\bf{\Lambda}},{\bf{y}},{\bf{k}})\propto
\sum_{\substack{Q,P \in \cr S_{N+M}}}\sum_{R\in S_M}[P|Q|R]\exp\left(\imath \sum_{n=1}^{N+M}
k_{Pn}X_{Qn}\right)\prod_{n=0}^{N+M}\Theta(X_{Q(n+1)}-X_{Qn})
  \nonumber \  ,
\end{eqnarray}
where the set ${\bf X } = \{{\bf x},{\bf y}\}$ comprises all coordinates 
and $X_{Q0}=-\infty$ and $X_{Q(N+M+1)}=+\infty$. A sector,
that is an ordering of particles, is labeled by the
permutations $Q$. The ordering of
particles corresponding to a permutation $Q$ is given by
\begin{eqnarray}
-\infty <X_{Q1}< X_{Q2} < \cdots <
X_{Q(N+M)}<+\infty \ .
\end{eqnarray}
In each sector the wave function \eqref{eq1.0-1} is a
superposition of plane waves where the coefficients $[P|Q|R]$ are
coordinate independent within a sector.
They are usually written as
follows (see e.g. \cite{tak99}): Let all the $N+M$ particles be ordered
and the spin--up particles be located at the integer positions 
\begin{eqnarray}
 1\leq f_1<f_2< \ldots < f_M\leq (N+M) \ .\label{eq2.6}
\end{eqnarray}
Then solution for the amplitudes $[P|Q|R]$ can
 be cast into the from 
\begin{eqnarray}
 [P|Q|R]=\sgn(R)\prod_{j<l}^M(\Lambda_{Rj}-\Lambda_{Rl} -\imath 2c)
\prod_{j=1}^MF_P(f_j,\Lambda_{Rj}),\label{eq2.6-1}
\end{eqnarray}
where
\begin{eqnarray}
 F_P(f,\Lambda)=\prod_{i=1}^{f-1}
((k_{Pi}-\Lambda+\imath c)
\prod_{l=f+1}^{N+M}((k_{Pl}-\Lambda-\imath c) \ .\label{eq2.7}
\end{eqnarray}
Thus the full wave function is a sum over the product of three permutation groups 
$S_{N+M}^{\times 2}\times S_M$.
Although the Ansatz \eqref{eq1.0-1} is relatively simple the full wave
functions turns out to be a rather cumbersome object due
to the involved structure of the amplitudes \eqref{eq2.6-1} and 
the summations over the permutation groups in Eq.~\eqref{eq1.0-1}.

\section{Eigenfunctions as Determinants}\label{sec2.1}

We cast Eq.~\eqref{eq1.0-1} in
a determinantal form with is particularly suited for the case when the thermodynamic limit is taken only for one species.  For the simplest case $M=1$ this has been
achieved in \cite{rech11}.  Here we state the generalization of this
result:
\begin{theorem}\label{theo1}
 The eigenfunctions of the Hamiltonian in Eq.~\eqref{eq1.0-1} can be cast into
    the form
    \begin{eqnarray}
    \label{eq1.1} 
    \Psi({{\bf x}},{{\bf k}},{{\bf y}}, {\bf {\Lambda}})&\propto& \sum_{R \in S_M}
    \sgn(R)\prod_{j<l}^{M}[
    \imath (\Lambda_{R j} -\Lambda_{R l}) +2c \sgn(y_l-y_j)]
    \Phi({\bf x},{\bf k},{\bf y,\Lambda}) \ ,
    \end{eqnarray}
where $\Phi({\bf x},{\bf k},{\bf y,\Lambda})$ is  given by
    the $(N+M)\times(N+M)$ determinant 
    \begin{eqnarray}
    \label{eq1.2}
 \Phi({\bf x},{\bf k},{\bf y,\Lambda})&=&
  \det\left[\prod_{s=1}^M
    A_j(\Lambda_{Rs},x_l-y_s)e^{\imath
    k_jx_l}\right| 
\left.\prod_{s\neq m}^M
    A_j(\Lambda_{Rs},y_m-y_s)e^{\imath
    k_jy_m} \right]_{\substack{j=1,\ldots,N+M \cr l=1,\ldots,N
 \cr m=1,\ldots,M}}
    \end{eqnarray}
    and
    \begin{eqnarray}
    A_j(\Lambda,x)=\imath(k_j -\Lambda) +c\sgn(x) \ .\label{eq1.3}
    \end{eqnarray}
%\item For periodic boundary conditions the normalization constant is
%      given by ?
%      \begin{eqnarray}
%      \fl |C|^{-2}=(N+M)!L^{N+M}\prod_{j=1}^{N+M}\left((k_j-\Lambda_m)^2 +c^2 +
%      \frac{2c}{L}\right)\\
%           \hspace*{6cm} \sum_{j=1}^{N+M}\frac{1}{(k_j-\Lambda_m)^2 +c^2 +2c/L} \nonumber .
%      \end{eqnarray}
The wave functions \eqref{eq1.1} are eigenfunctions of the Hamiltonian
      \eqref{eq1.0-1} and of the center of mass momentum operator
      \begin{eqnarray}
       \hat{K}=\frac{1}{\imath}\left(\sum_{n=1}^{N}\frac{\partial}{\partial x_n}
    +\sum_{m=1}^M\frac{\partial}{\partial y_m}\right) \label{eq1.3-1}
      \end{eqnarray}
      to the eigenvalues
       $E=$ $\sum_{j=1}^{N+M} k_j^2$ and $K=$ $\sum_{j=1}^{N+M}k_j$.    
       \end{theorem}
We prove Theorem \ref{theo1} in App.~\ref{appA}.
Comparing Eq.~\eqref{eq1.1} with the original form  \eqref{eq1.0-1} shows that essentially 
the summations over two of the three permutation groups were replaced by a
$(N+M)\times(N+M)$ determinant. This is at least in principle more
convenient than Eq.~\eqref{eq1.0-1} since it allows to
employ the powerful methods of matrix algebra to manipulate
determinants.

We briefly discuss the symmetries of the functions $\Psi({\bf x},{\bf k},{\bf y},{\bf\Lambda})$ in \eqref{eq1.1}. Due to the determinantal form of $\Phi({\bf x},{\bf k},{\bf y},{\bf\Lambda})$ 
the antisymmetry of   $\Psi({\bf x},{\bf k},{\bf y},{\bf\Lambda})$ in ${\bf x}$ and in ${\bf k}$ 
is obvious. To show that $\Psi$ is antisymmetric in ${\bf y}$ we act with an arbitrary permutation $P$ on ${\bf y}$ as $P({\bf y}): $ $y_{\mu}\leftrightarrow y_{P\mu}$ 
and write Eq.~\eqref{eq1.1} as
 \begin{eqnarray}
\Psi({\bf x},{\bf k},P({\bf y}),{\bf \Lambda})&=&\sum_{R\in S_M} \sgn(P)\sgn(R)
\Phi({\bf x},{\bf k}, P({\bf y}),P({\bf \Lambda}))\nonumber\\
&&\qquad \prod_{\substack{j<l}}^M\left[\imath(\Lambda_{PRj}-\Lambda_{PRl})+2c\ \sgn(y_{Pl}-y_{Pj})\right] \label{eq1.5-1}
\end{eqnarray}
We observe that  $\Phi$ as well as the second line in Eq.~(\ref{eq1.5-1}) 
are antisymmetric under the simultaneous action of the permutation $P({\bf y})$ and  $P({\bf \Lambda}):$ $ \Lambda_{R\mu}\leftrightarrow \Lambda_{PR\mu}$. Thus a minus sign is 
picked up by the sign of the permutation  $\sgn(P)=-1$. The antisymmetry $\Psi({\bf x},{\bf k},{\bf y},P({\bf \Lambda}))$ $= -\Psi({\bf x},{\bf k},{\bf y},{\bf \Lambda})$ can be proven similarly. 
Finally we note that the wave function \eqref{eq1.1}
has no well defined symmetry when a spin-up and
a spin-down particle are exchanged.

Imposing periodic boundary conditions  on \eqref{eq1.5-1}  yields a set of coupled algebraic equations, which are known as  Bethe Ansatz equations  \cite{yan67}
\begin{eqnarray}
 k_jL &=&  2\pi n_j -2\sum_{m=1}^{M}\arctan\left(\frac{k_j-\Lambda_m}{c}\right)\quad ,\quad j=1,\ldots,N+M \ ,\label{eq2.10}\\
 2\pi J_{\mu}& =& 2\sum_{\substack{j=1}}^{N+M}\arctan\left(\frac{k_{j}-\Lambda_{\mu}}{c}\right)
+2\sum_{\substack{\nu=1\cr \neq
\mu}}^{M}\arctan\left(\frac{\Lambda_{\mu}-\Lambda_{\nu}}{2c}\right)
~ ,~ \mu =1,\ldots,M \ .  \label{eq2.11}
\end{eqnarray}
The quantum numbers $J_{\mu}$ are
integers for $N$ odd and half-odd integers for $N$ even. The quantum numbers $n_j$ are
integers for $M$ even and half-odd integers for $M$ odd. For
convenience we will always assume in the following $N$ to be odd and $M<N$.
The values for $J_{\mu}$ are bounded by
\begin{eqnarray}
-\frac{N+2M-1}{2}\leq J_{\mu}\leq\frac{N+2M-1}{2}\label{eq2.12} \ .
\end{eqnarray}
In the ground state  the $n_j$ are adjacent integers or half odd integers ranging from $(N+M-1)/2$ to   $-(N+M-1)/2$ and for an odd number of spin--up particles the $ J_{\mu} $ are  chosen as 
\begin{eqnarray}
% \left\{n_j\right\}_{j=1,\ldots,N+M}&=& \left\{ 0,\pm 1,\ldots, \pm \frac{N+M-1}{2}\right\} \label{eq2.14-1} \ ,\\
 \left\{J_{\mu}\right\}_{\mu=1,\ldots,M}&=&
\left\{ 0,\pm 1,\ldots, \pm \left(\frac{M}{2}-1\right),\frac{M}{2}\right\}
\label{eq2.14-2} \ .
\end{eqnarray}
 In the following we will assume that for $c=0$ the spin up particles have free momenta 
 $-k_{\uparrow \mu}$, which lie in the interval $- k _{\rm F} < k_{\uparrow \mu} < k_{\rm F}$, where
$k_{\rm F} =\pi N/L$ is the Fermi momentum of the non--interacting spin down particles.
 In this range the quantum numbers $J_{\mu}$
can be identified with the momentum of the non--interacting spin up particle. 
They indicate the single particle states, which for $c\rightarrow 0^+$ become doubly occupied. This means in the limit $c\to 0$: 
$\Lambda_\mu(c)\to -k_{\uparrow\mu}$ and $k_{\uparrow \mu}=2\pi J_\mu/L$. 
%They have the dimension of a momentum and
%vary between $\pm k_{\rm max}$  as $J_{\mu}$ varies between
%the bounds given in Eq.~\eqref{eq2.12}.
For the systems overall momentum $K$ follows from Eqs.~\eqref{eq2.10} and \eqref{eq2.11}
\begin{eqnarray}
K=\sum_{j=1}^{N+M}k_j = \frac{2\pi}{L}\left(\sum_{j=1}^{N+M} n_j
-\sum_{\mu=1}^M J_\mu\right) \label{eq2.17} \ .
\end{eqnarray}
If the set of quantum numbers $n_j$ is determined by the ground state the first sum vanishes
and  $K=-\sum_{\mu=1}^M k_{\uparrow\mu}$. 
%In the 
%lab frame where the spin-down Fermions are at rest the 
%latter is identified with the center of mass momentum of 
%the spin-up particles.  

\section{Thermodynamic limit}\label{sec2.3}

In a highly imbalanced system the density of the minority Fermions is very low. Thus 
one can assume that their thermodynamics is well approximated by a virial expansion.

In this approach the thermodynamical  limit is taken only for
the spin--down particles. The number of spin--up Fermions is finite.  Since all spin--up particles have free momenta 
smaller than $k_{\rm F}$, the quantum numbers $n_j$ are those of the ground state.
The density of the quasi-momenta is obtained by taking the
derivative of Eq.~\eqref{eq2.10}
\begin{eqnarray}
\varrho(k)=\frac{L}{2\pi}
+\sum_{\mu =1}^M\frac{1}{\pi}\frac{c}{(k-\Lambda_\mu)^2 +c^2}
\label{eq2.18} \ .
\end{eqnarray}
%Equation \eqref{eq2.18} has a rather natural interpretation: The first term
%corresponds to the density of states of a non-interacting
%Fermi-sea formed by the spin-down particles. The $M$ additional
%terms appear due to the presence of the spin-up Fermions. Each of
%them has the form of a Lorentzian-distribution centered around
% $\Lambda_l$. For $c\rightarrow 0^+$ they yield
%$\delta$-functions $\delta(k-\Lambda_l)$.
This is in leading order in $L$ to the momentum distribution of a sea of free Fermions. 
The quasi-momenta distribute themselves with the constant density $L/(2\pi)$ between two values $ k_{\pm}$. 
The $M$ additional terms on the right hand side of Eq.~\eqref{eq2.18} then might be 
 interpreted as momentum distribution for 
the spin-up particles. The momenta $k_{\pm}$ are 
defined through the conditions 
\begin{eqnarray}
N+M&\stackrel{!}{=}&\int\limits_{k_{-}}^{k_{+}}dk \varrho(k) \nonumber\\
K &\stackrel{!}{=}&\int\limits_{k_{-}}^{k_{+}}dk k \varrho(k) 
\label{eq2.18-1} 
\end{eqnarray}
which are transcendental equations. Assuming that the non--interacting Fermi sea is at rest, the solutions can be expanded 
in inverse powers of the system size $L$
\begin{eqnarray} 
\label{eq2.18-2} 
 k_{\pm} &= & \pm k_{\rm F} \pm \frac{2}{L}\sum_{\mu=1}^M
\left(  \frac{\pi}{2} - \arctan \left(\frac{k_{\rm F}\mp \Lambda_\mu}{c}\right)\right)+{\cal O}(L^{-2}) 
%\fl &&\qquad\quad \times 
%\left[1-\frac{1}{L}\frac{c}{(k_{\rm F}-\Lambda_\mu)^2+c^2}-\frac{1}{L}\frac{c}{(k_{\rm F}+\Lambda_\mu)^2+c^2}\right]+ {\cal O}(L^{-3}) \ ,
\end{eqnarray}
where  we introduce
 the notation
\begin{eqnarray}
\label{eq2.19a}
 v(\Lambda, a)&=&\frac{1}{\pi}\int\limits_{-a}^a dk
\arctan\left(\frac{k- \Lambda}{c}\right).
\end{eqnarray}
Using Eq.~\eqref{eq2.18} the second
Bethe-Ansatz equation \eqref{eq2.11} can be written as
\begin{eqnarray}
\label{eq2.19}
k_{\uparrow\mu} &=&v(\Lambda_{\mu},k_{\rm F})+ w_1(\Lambda_{\mu})+ \sum_{\substack{\nu=1\cr \neq \mu}}^{M} w_2(\Lambda_{\mu},\Lambda_\nu)\ , 
\end{eqnarray}
where the functions $w_1(\Lambda_{\mu})$ and $w_2(\Lambda_{\mu},\Lambda_\nu)$ scale ${\cal O}(L^{-1})$ with the system size
\begin{eqnarray}
\label{eq2.19b}
w_1(\Lambda_{\mu})&=&\frac{\pi}{L}\frac{\partial v(\Lambda_\mu,k_{\rm F})}{\partial k_{\rm F}}\nonumber\\
w_2(\Lambda_{\mu},\Lambda_\nu)&=& \frac{2c}{L\pi}\int\limits_{-k_{\rm F}}^{+k_{\rm F}}dk
\frac{ \arctan\left(\frac{k- \Lambda_{\mu}}{c}\right)}{(k-\Lambda_\nu)^2 +c^2} +\frac{2}{L}\arctan\left(\frac{\Lambda_{\mu}-\Lambda_{\nu}}{2c}\right)\ .
\end{eqnarray}
Thus in leading order in $L$ the quantity $\Lambda_\mu$ is determined by $\hat{J}_\mu$ only. 
Terms that couple different $\Lambda_{\mu}$'s are of order $1/L$. For $c>0$ the function $v$ in (\ref{eq2.19a}) is monotonously decreasing in $\Lambda$. Thus it can be inverted.  To this end the identity $\arctan(x)$$=$ $\sgn(x)\pi/2$ $-\arctan(x^{-1})$ is plugged into equation (\ref{eq2.19a}). The first term containing the sign--function can be integrated. One obtains an implicit equation for  $\Lambda_\mu$ which can be iterated. The inverted function reads  to first order in the coupling strength
 \begin{eqnarray}
 \Lambda_\mu &\approx&- k_{\uparrow\mu} -\frac{c}{\pi}\ln\left|\frac{k_{\rm F}+k_{\uparrow\mu}}{k_{\rm F}-k_{\uparrow\mu}}\right|+{\cal O}(c^2) \ .
 \end{eqnarray}
For $c\to\infty$ the quantity $\Lambda_\mu$ scales as  $\Lambda_\mu = c \lambda_\mu$. The integral (\ref{eq2.19a}) becomes trivial and $\Lambda_\mu$$=$ $-c \tan( \pi k_{\uparrow\mu}/2k_{\rm F})$. 
The total energy is given by
\begin{eqnarray}
 E&=&\int\limits_{k_-}^{k_+}dk k^2 \varrho(k)\ =\  \frac{L (k_{+}^3-k_-^3)}{6 \pi}+\frac{1}{\pi}\sum_{\mu=1}^M\int
\limits_{k_-}^{k_+}dk
\frac{c k^2}{(k-\Lambda_\mu)^2 +c^2}\label{eq2.20} \ ,
\end{eqnarray}
This equation can be expanded in inverse powers of the system size using equation (\ref{eq2.18-2}). All terms ${\cal O}(L^{-2})$ are neglected.  
The standard integral on the right hand side of Eq.~\eqref{eq2.20} can be evaluated and the total energy can be written as
\begin{eqnarray}
E &=& E_{\rm F} +  \sum_{\mu=1}^M E^{(1)}(\Lambda_\mu) +  \sum_{\mu < \nu}^M E^{(2)}(\Lambda_\mu,\Lambda_\nu)\label{energy} \ ,
\end{eqnarray}
where $E_{\rm F}=Lk_{{\rm F}}^3/(3\pi)$ denotes the energy of the non--interacting sea particles. The single particle energy  $E^{(1)}(\Lambda_\mu)$ reads in leading order 
\begin{eqnarray}
\label{eq2.22}
E^{(1)}(\Lambda_\mu) &=&  k_{\rm F}^2 +\frac{c}{\pi}\left[2k_{\rm F} +\Lambda_\mu\ln\left(\frac{c^2+(\Lambda_\mu- k_{\rm F})^2}
{c^2+(\Lambda_\mu+ k_{\rm F})^2}\right)\right] \\
&&+ \left(k_{\rm{F}}^2-\Lambda_\mu^2 +c^2\right)\frac{\partial v(\Lambda_\mu,k_{\rm F})}{\partial \Lambda_\mu}+{\cal O}(L^{-1})\nonumber \ .
\end{eqnarray}
The two particle energy can be expanded in the same way. Here the leading order term is ${\cal O}(L^{-1})$
\begin{eqnarray}
\label{eq2.22a}
  E^{(2)}(\Lambda_\mu, \Lambda_\nu)  & = &   
        \sum _{\sigma= \pm 1} \frac{ 2 k_{\rm F}}{\pi L}\left(2-k_{\rm F}\frac{d}{d k_{\rm F}}\right)\\
 &&   \left[\frac{\pi}{2} - \arctan\left(\frac{k_{\rm F}+\sigma \Lambda_\mu}{c}\right)\right] 
   \left[\frac{\pi}{2} - \arctan\left(\frac{k_{\rm F}+\sigma \Lambda_\nu}{c}\right)\right]
   +{\cal O}(L^{-2}) \  \nonumber .
\end{eqnarray}
%The first term on the right hand side scales like $L$ whereas the 
%remaining $M$ contributions are of the order one. Therefore when
%substituted into the the equation above the corrections to $k_{\rm F}^3$ 
%up to order $1/L$ need to be taken into account. From 
%Eq.~\eqref{eq2.18-1} we find for contribution due to 
% the first term in Eq.~\eqref{eq2.20}
%\begin{eqnarray}
%E&=& E_{\rm F} +\frac{k_{\rm F}^2}{\pi}\sum_{\mu =1}^M \left[\pi -\arctan\left(\frac{k_{\rm F} -\Lambda_{\mu}}{c}\right)
%-\arctan\left(\frac{k_{\rm F} +\Lambda_{\mu}}{c}\right)\right] \label{eq2.20-1} \ .
%\end{eqnarray}
Since the energy of the  non-interacting system is $E_0$$=$$E_{\rm F}+$$\sum_{\mu=1}^M k_{\uparrow\mu}^2$ the total energy
shift $\Delta E_M $ = $E-E_0$ for $M$ spin--up particles can in leading order in the system size be expressed as
\begin{eqnarray}
\Delta E_M({\bf k}_\uparrow) &=& \sum_{\mu=1}^M\left( E^{(1)}(\Lambda_\mu) - k_{\uparrow\mu}^2 \right)\label{eq2.23} \ .
\end{eqnarray}
where $\Lambda_{\mu}$ is determined 
by $k_{\uparrow\mu}$ only. The energy shift is 
additive,  that is each contribution corresponds to the energy shift
caused by the interaction of a single spin-up particle 
with the Fermi-sea. In subleading order $\Lambda_{\mu}$ $=\Lambda_{\mu}({\bf k}_\uparrow)$ depends on all initial momenta $k_{\uparrow\mu}$, $\mu=1,\ldots, M$.

\section{Effective interaction of two spin-up Fermions}\label{sec3}

\begin{figure}[h]
\begin{center}
\include{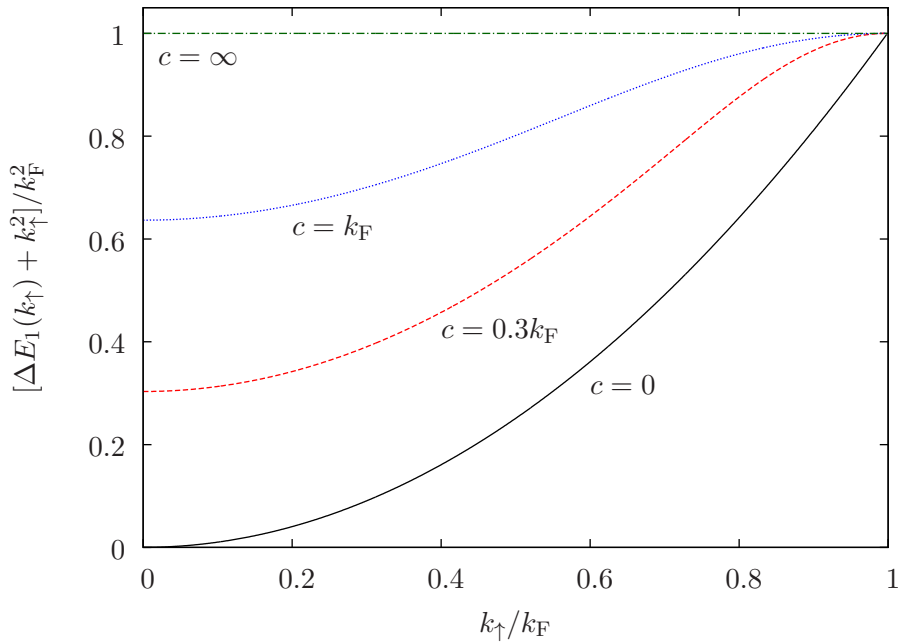}
\caption{\label{Fig:Dispersion} Single particle dispersion $E^{(1)}(k_{\uparrow})$ for different values of the interaction strength $c$. The values are $c = 0$ solid line (black),  $c=0.3k_{\rm F}$ dashed line (red),  $c= k_{\rm F}$  dotted line (blue) and $c = \infty$ dashed--dotted line (green). }
\label{fig1}
\end{center}
\end{figure}

The single particle energy $E^{(1)}(\Lambda_\mu)$ is additive as a function of $\Lambda_\mu$, however it is not additive as a function of the free momenta $k_{\uparrow\mu}$. 
Thus the functional form of the total energy shift changes when a spin--up particle is added to the system.   
The total energy shift of a system of $M$ spin--up particles $\Delta E_M$ 
can be expanded in a cluster expansion as  
\begin{eqnarray}
\Delta E_1(k_{\uparrow 1})&=& \left.E^{(1)}(\Lambda_1)\right|_{\Lambda_1=v^{-1}(k_{\uparrow 1})}- 
k_{\uparrow 1}^2\nonumber\\
 \Delta E_2(k_{\uparrow 1}, k_{\uparrow 2})&=&\Delta E_1(k_{\uparrow 1})+\Delta E_1(k_{\uparrow 2}) + W_2(k_{\uparrow 1},k_{\uparrow 2}) \nonumber\\
 \Delta E_3(k_{\uparrow 1}, k_{\uparrow 2}, k_{\uparrow 3})&=&\sum_{\mu=1}^3\Delta E_1(k_{\uparrow\mu})+ \sum_{\mu> \nu}^3 W_2(k_{\uparrow\mu}, k_{\uparrow\nu})+ W_3(k_{\uparrow 1},k_{\uparrow 2},k_{\uparrow 3})  \nonumber\\
   &\vdots&\qquad\qquad\  \label{eq3.0}
\end{eqnarray}
In the following we focus  on the first two terms in this expansion, 
which contain the information about the single particle 
dispersion and  on the interaction between two spin--up particles in the presence of the Fermi--sea.  The single particle energy $E^{(1)}(k_\uparrow)$ given in equation \eqref{eq2.22} is plotted in figure 
\ref{Fig:Dispersion}. 
It is seen that the dispersion relation is at least for small values of $k_{\uparrow}$ still approximately quadratic $E(k) \approx $ $\Sigma + k^2/m^*$. Self--energy and renormalized mass are given in terms of $a= c/k_{\rm F}$ by 
\begin{eqnarray}
\Sigma &=& k_{\rm F}^2 \left( \frac{2 a}{\pi} -a^2 + \frac{2}{\pi}\left(1+a^2\right)\arctan a \right)\nonumber\\
m^*&=&\frac{\pi/2 - 2\arctan a + (2/\pi)\arctan^2 a}{\pi/2 -\arctan a - a/(1+a^2)}
\end{eqnarray}
For finite interaction strength $c>0$ and for larger momenta $k_\uparrow$ the dispersion relation deviates from the quadratic behavior, for $k_\uparrow= k_{\rm F}$ the first derivative vanishes, leading to a van Hove like singularity in the density of states. 
Formally the Fermi momentum of the majority Fermions $k_{\rm F}$ plays the role of an inverse lattice spacing as momentum cutoff. For $c=\infty$ the mass becomes formally infinite and the energy momentum independent.

The effective interaction energy  $ W_2(k_{\uparrow 1}, k_{\uparrow 2})$ is a function of $k_{\uparrow 1}$ and  $k_{\uparrow 2}$ rather than of $\Lambda_1$ and  $\Lambda_2$.  
Therefore it is not just given by  the sum of the two--particle energies $E^{(2)}(\Lambda_1,\Lambda_2)$ and $E^{(2)}(\Lambda_2,\Lambda_1)$ but also the 
single particle energies $E^{(1)}(\Lambda_\mu)$ contribute.  Using Eqs.~\eqref{eq2.19} and \eqref{eq2.19b} one finds
\begin{eqnarray}
 W_2(k_{\uparrow 1},k_{\uparrow 2})&=& - \frac{\partial E^{(1)}(\Lambda_1)}{\partial \Lambda_1}\left(\frac{\partial v(\Lambda_1, k_{\rm F})}{\partial \Lambda_1}\right)^{-1} w_2(\Lambda_1,\Lambda_2) \label{eq3.6}\nonumber\\
 &&- \left. \frac{\partial E^{(1)}(\Lambda_2)}{\partial \Lambda_2}\left(\frac{\partial v(\Lambda_2, k_{\rm F})}{\partial \Lambda_2}\right)^{-1} w_2(\Lambda_2,\Lambda_1) + E^{(2)}(\Lambda_1,\Lambda_2)
\right|_{\substack{\Lambda_1=v^{-1}(k_{\uparrow 1})\cr \Lambda_2=v^{-1}(k_{\uparrow 2})}} \ .
\end{eqnarray}
which has for general $c>0$ to be treated numerically, since the function $v(\Lambda,a)$ in Eq.~(\ref{eq2.19a}) can only be inverted numerically. 
An expansion of Eq.~\eqref{eq3.6} for small interaction strength is possible and yields for $k_{\uparrow 1}<k_{\rm F}$ and  $k_{\uparrow 2}<k_{\rm F}$ 
\begin{eqnarray}
 \lim_{c\rightarrow 0^+}W_2(k_{\uparrow 1},k_{\uparrow 2})&=&
-\frac{4 c k_{\uparrow 1} k_{\uparrow 2}}{L}\left(\frac{1}{k_{{\rm F}}^2-
k_{\uparrow 1}^2} + \frac{1}{k_{{\rm F}}^2-k_{\uparrow 2}^2}\right) \ .
 \ \label{eq3.10-1}
\end{eqnarray}
%If either $k_{\uparrow 1}$ or $k_{\uparrow 2}$ approaches $k_{\rm F}$ a singular contribution has to be added in Eq.~(\ref{eq3.10-1}). These points are therefore excluded. 
This expression becomes singular if either of the two momenta approaches the Fermi momentum.
An asymptotic expansion of Eq.~\eqref{eq3.6} for strong interaction yields
\begin{eqnarray}
W_2(k_{\uparrow 1},k_{\uparrow 2})&=& \frac{2\pi k_{\rm F}}{L} \left\{1+ \frac{k_{\uparrow 1} k_{\uparrow 2}}{k_{\rm F}^2}- \frac{2 k_{\rm F}}{c \pi}\left[\cos^2\left(\frac{\pi k_{1\uparrow}}{2k_{\rm F}}\right) + \cos^2\left(\frac{\pi k_{1\uparrow}}{2k_{\rm F}}\right)\right]\right\}+ {\cal O}(c^{-2})
\label{eq3.10} \ .
\end{eqnarray}
Figures~\ref{fig1} show the plots of the interaction potential as a function of $a= c/k_{\rm F}$ for zero center--of--mass momentum and for different relative momentum 
$k$$=k_{1\uparrow}$$=-k_{2\uparrow}$ . The curves depend crucially on the relative momentum $k$. Whereas for small momentum the interaction energy 
increases monotonously with $c$, for higher momenta the interaction energy has a maximum for small interaction strength $c$ and 
decays for strong coupling strength to a value given by equation (\ref{eq3.10}). Only for $k= k_{\rm F}$ it decays to zero.  

The interaction energy depends non--trivially on both arguments. This means that translation 
invariance of the reduced system of minority Fermions is broken by the Fermi sea. The dependence on the total momentum becomes most striking in the small coupling limit  (\ref{eq3.10-1}). 
For $k_{\uparrow 1}k_{\uparrow 2}<0$ the interaction energy 
$W_2(k_{\uparrow 1},k_{\uparrow 2})$  is always
positive but for $k_{\uparrow 1}k_{\uparrow 2}>0$ and for
small values of $c$ it becomes negative. This can be seen  in figure \ref{fig1b}, where  $W_2(k_{\uparrow 1},k_{\uparrow 2})$ is plotted for zero relative momentum and different values of the center of mass momentum. 
Although the dependence of the interaction on the center of mass momentum is certainly an interesting feature, we focus in the sequel mainly on $K=0$.
%%%%%%%%%%%%%%%%%%%%%%%%%%%%%%%%%%
\begin{figure}[h]
\begin{center}
\include{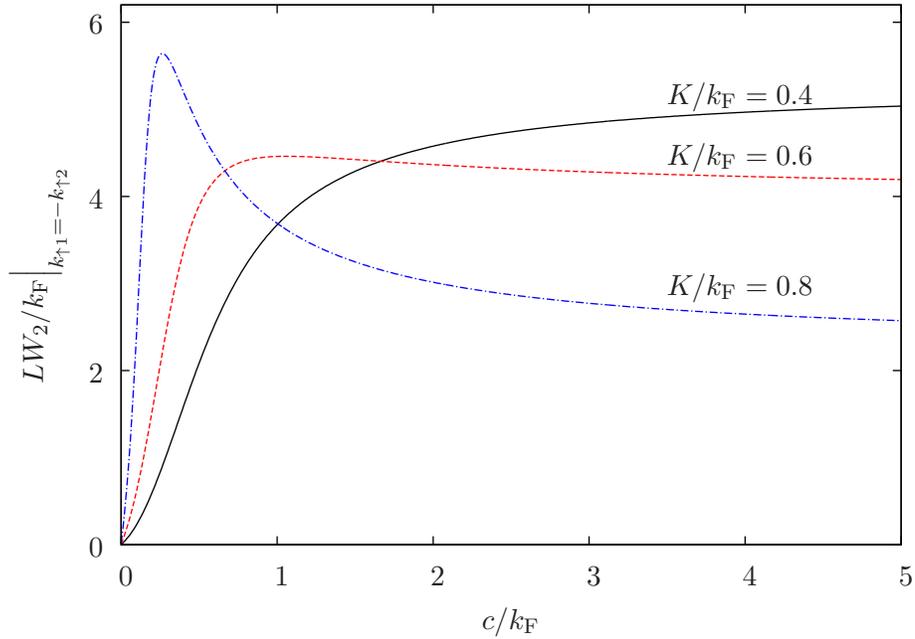}
\caption{Interaction energy--shift  $W_2(k_{\uparrow 1},k_{\uparrow 2})$ for zero center of mass momentum $ K= 0$ for different values of the free momenta $k_{\uparrow 1} =$ $ - k_{\uparrow 2}$. The values are:  $k_{\uparrow 1}=0.4$ solid
line (black), $k_{\uparrow 1}=0.6$ dashed
line (red), $k_{\uparrow 1}=0.8$ dot-dash
line (blue) . }
\label{fig1a}
\end{center}
\end{figure}
%%%%%%%%%%%%%%%%%%%%%%%%%%%%%%%%%%%%%%%
\begin{figure}[h]
\begin{center}
\include{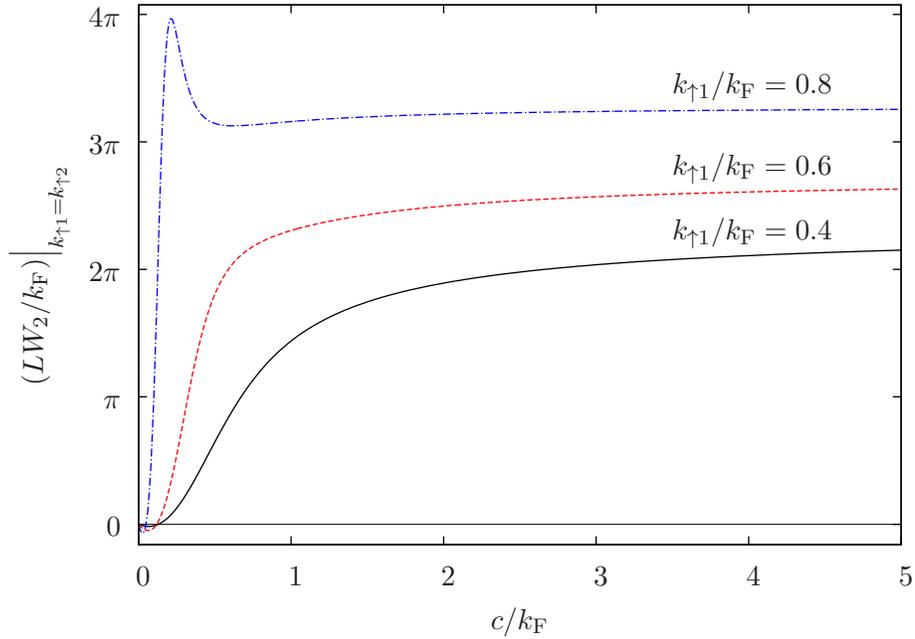}
\caption{Interaction energy--shift  $W_2(k_{\uparrow 1},k_{\uparrow 2})$ for zero relative momentum and for different values of the center of mass momentum $K =0.4$ solid
line (black), $K=0.6$ dashed
line (red), $K=0.8$ dot-dash
line (blue) . }
\label{fig1b}
\end{center}
\end{figure}
%%%%%%%%%%%%%%%%%%%%%%%%%%%%%%%%%%%%%%

Neglecting the higher order interactions in the cluster expansion \eref{eq3.0} the thermodynamical limit can now be taken for the spin--up particles as well.  
In the ground state the $M$ $<N$ quantum numbers $J_n$ are adjacent integers distributed around zero. We introduce an imbalance parameter $\eta$ $=M/N$, 
which varies between zero  and one for the balanced case. The ground state energy per unit length becomes 
\begin{eqnarray}
\label{energ}
 \frac{E_0}{L}&=& \frac{k_{{\rm F}}^3}{3\pi} +  \frac{1}{2\pi} 
\int\limits_{-\eta k_{\rm F}}^{\eta k_{\rm F}}E^{(1)}(\Lambda(x)) dx 
+ \frac{L}{8\pi^2} \int\limits_{-\eta k_{\rm F}}^{\eta k_{\rm F}} dx 
\int\limits_{-\eta k_{\rm F}}^{\eta k_{\rm F}} dx^\prime E^{(2)}(\Lambda(x),\Lambda(x^\prime))\ .
\end{eqnarray}
It can be checked that the expansion of this expression in powers of the imbalance parameter $
\eta$ 
% \begin{eqnarray}
%  \frac{E_0}{L} &=& \frac{k_{{\rm F}}^3}{3\pi}+ \frac{\eta k_{\rm F}}{\pi}\left(k_{\rm F}^2+\frac{2 c k_
%{\rm F}}{\pi} -\frac{2}{\pi} \left(c^2+k_{\rm F}^2\right)\arctan\left(\frac{k_{\rm F}}{c}\right) \right)+ 
%  \frac{\eta^2 k^3_{\rm F}}{\pi}\left(1-\frac{2}{\pi} \arctan\left(\frac{k_{\rm F}}{c}\right)\right)^2  
% \end{eqnarray}
 coincides up to second order with the result of Suzuki \cite{suz69}. The energies of the second 
order cluster expansion \eqref{eq3.0} constitute the spectrum of an effective few body--
Hamiltonian for the spin--up particles
\begin{eqnarray}
\label{Heff}
\hat{H}_{\uparrow\uparrow} &=& \sum\limits_{|k|<k_{\rm F}} \left(k^2+\Delta E_1(k)\right) \hat{c}_{\uparrow k}^\dagger
\hat{c}_{\uparrow k}+\sum\limits_{|k|<k_{\rm F}} \sum\limits_{|k^\prime|<k_{\rm F}}  
\hat{c}_{\uparrow k}^\dagger\hat{c}_{\uparrow k} W_2(k,k^\prime) \hat{c}_{\uparrow k^\prime}^
\dagger\hat{c}_{\uparrow k^\prime} 
\end{eqnarray}  
where the  anticommuting operators $\hat{c}_{\uparrow k}^\dagger$ ($\hat{c}_{\uparrow k}$) 
create (annihilate) a spin--up polaron with free momentum $k$. 
The interaction energy 
$W_2(k_1,k_2)$ can be related to an effective interaction potential of the minority Fermions via
\begin{eqnarray}
W_2(k_{1\uparrow},k_{2\uparrow})&=& \int\limits_0^L dy_1   \int\limits_0^L dy_2 V_{\uparrow\uparrow}
( y_1,y_2) R_0(y_1,y_2;k_{1\uparrow},k_{2\uparrow}) \ ,
\end{eqnarray}
where $R_0(y_1,y_2; k_{1\uparrow},k_{2\uparrow})$ is the density--density correlation function of 
two minority Fermions. Its precise definition 
is given in equation (\ref{eq4.0a}) in the next section, where it is calculated exactly for arbitrary 
momenta $k_{1\uparrow}$, $k_{2\uparrow}$ and interaction strength $c$.  
The simplest approximation for  the interaction potential, which is essentially  a  Born 
approximation, is obtained by replacing $R_0$ by its non--interacting value
\begin{eqnarray}
R_0(y_1,y_2; k_{1\uparrow},k_{2\uparrow})&=&\frac{2}{L^2}\left( 1-\cos\left((k_{\uparrow 1}-k_
{\uparrow 
2})(y_1-y_2)\right) \right)\label{eq4.24} \ .
\end{eqnarray}
Focussing on $K=0$ implies 
 $V_{\uparrow \uparrow}( y_1,y_2)$$= V_{\uparrow \uparrow}( y_1 -y_2)$. 
 Thus for small coupling $c$ the effective interaction is essentially  the Fourier transform of the 
interaction energy. Using the small $c$ expansion (\ref{eq3.10-1}) of the interaction energy the 
potential is given by
 \begin{eqnarray}
 \label{pot}
 V(x)&=& c \delta(x) + c k_{\rm F} \sin (2k_{\rm F} x)
\end{eqnarray} 
The Hamilton--operator  (\ref{Heff}) is bounded.  The Fermi momentum acts as a momentum cutoff. 
Thus also the interaction potential can be determined only up to a length scale of order 
$k_{\rm F}$. This uncertainty is can into account by convolution with a proper distribution of width 
$k_{\rm F}^{-1}$. The oscillatory term in the potential (\ref{pot}) cancels.  Choosing for definiteness 
the characteristic function $\chi_{[-k_{\rm F},k_{\rm F}]}$ for the convolution, the interaction 
between two minority Fermions is zero everywhere but for distances smaller than $k_{\rm F}$, 
where it is constant $V(x)= 2 c k_{\rm F}$.        

\section{Density-density correlation functions}\label{sec3.2}

Using the determinantal representation (\ref{eq1.1}) of the many--body wave function the density-density correlation function of two minority Fermions and even the three point correlation function of two minority Fermions and one majority Fermion can be calculated exactly.  The latter yields insight to what extent two impurities affect the otherwiese flat density profile of the Fermi sea. 

The general density-density correlation function for
two spin-up and $n$-spin-down Fermions is defined in
coordinate representation through the multiple integral
\begin{eqnarray}
&&R_n(y_1,y_2,x_1,\ldots,x_n; k_{1\uparrow},k_{2\uparrow})\ := \ 4N^n\int\limits_0^Ldx_{n+1}\cdots\int\limits_0^Ldx_{N}\left|\Psi({\bf x},{\bf k}, y_1,y_2,\Lambda_1,\Lambda_2)\right|^2\label{eq4.0a} \ .
\end{eqnarray}
With the explicit form \eqref{eq1.1} of the eigenfunctions for
$M=2$ this can be written as
\begin{eqnarray}
R_n&\propto&\sum_{\substack{R,R^\prime \cr \in S_2}}\sgn(R+R^\prime)\left[\imath
(\Lambda_{R1} -\Lambda_{R2}) -2c\sgn(y_1-y_2)\right] \nonumber\\
&&\qquad\qquad\qquad\times \left[\imath ( \Lambda_{R^\prime2} -\Lambda_{R^\prime1})-2c\sgn(y_1-y_2)\right]\label{eq4.0}\\
&&\ \times\int\limits_0^Ldx_{n+1}\cdots\int\limits_0^Ldx_N \Phi({\bf
x}, y_1,y_2,\Lambda_{R1},\Lambda_{R2})\Phi^*({\bf x},
y_1,y_2,\Lambda_{R^\prime1},\Lambda_{R^\prime2})\ . \nonumber
\end{eqnarray}
Note that the quantities $\Lambda_{1,2}$ 
and therefore the correlation function itself depends on 
the momenta $k_{\uparrow 1}$, $k_{\uparrow 2}$ of the free spin--up particles.  We are interested in $n=0,1$. 
For both cases, the correlation function 
can be cast into the unified form.  We define $\tilde{R}_n$ $ = L^{2+n}R_n/4N^n$ and 
assume $y_1\leq y_2$ without loss of generality. Then $\tilde{R}_n$ can be written as
\begin{eqnarray}
\label{eq4.3-1} 
 \tilde{R}_n &=& 2^{-n}\mathcal{R}\left\{
\det[I^{(n)}(\Lambda_{1},\Lambda_{2})]-\re\left( e^{2i\arctan \left(\frac{\Lambda_1-\Lambda_2}{2c}\right)}
		   \det[J^{(n)}(\Lambda_{1},\Lambda_{2})]\right)\right\} \ ,
\end{eqnarray}
where the normalization constant $\mathcal{R}$ reads
\begin{eqnarray}
 \mathcal{R}^{-1}&=&\prod\limits_{j=1}^2\int\limits_{-k_{\rm F}}^{k_{\rm F}}dk\frac{k_{\rm F}}
{(k-\Lambda_j)^2+c^2} \label{eq4.18} -\re\left(\int\limits_{-k_{\rm F}}^{k_{\rm F}}dk\frac{ k_{\rm F} e^{i\arctan \left(\frac{\Lambda_1-\Lambda_2}{2 c}\right)} }
{[-\imath(k- \Lambda_1) + c][\imath(k-\Lambda_2)
+c]}\right)^2\ \nonumber
\end{eqnarray}
The quantities
$I^{(n)}(\Lambda_{1}, \Lambda_{2})$ $=$ $[I_{jl}(\Lambda_{1}, \Lambda_{2})]_{j,l=1,\ldots,n+2}$ and 
$J^{(n)}(\Lambda_{1}, \Lambda_{2})$ $=[J_{jl}(\Lambda_{1},\Lambda_{2})]_{j,l=1,\ldots,n+2}$  in
Eq.~\eqref{eq4.3-1} denote $(n+2)\times (n+2)$ matrices. Their
explicit form will be stated below. Beside its coordinate
dependence $\tilde{R}_n$
depends on the interaction strength $c$ and
$\Lambda_1$ and $\Lambda_2$. The latter two
quantities are determined by Eq.~\eqref{eq2.19}.
Details of the derivation of Eq.~\eqref{eq4.3-1} are presented in App.~\ref{appB}  In the following we treat the two cases $n=0,1$ 
separately.

\subsection{Two particle density-density correlation function}

The density--density correlation function of the two spin up particles 
corresponds to $n=0$ in the general expression \eqref{eq4.3-1}.
 We give the explicit form of the entries of the matrix
$I^{(0)}(\Lambda_{1},\Lambda_{2})$ and  $J^{(0)}(\Lambda_{1},\Lambda_{2})$. 
As shown in App.~\ref{appB} these are
\begin{eqnarray}
  I^{(0)}_{jj}(\Lambda_{1},\Lambda_{2}) &= & \int\limits_{-k_{\rm F}}^{+k_{\rm F}}\frac{ k_{\rm F} \, dk}{(k-\Lambda_{j})^2 +c^2} \nonumber \ , \qquad j=1,2\\
I^{(0)}_{12}(\Lambda_{1},\Lambda_{2}) &= & \int\limits_{-k_{\rm F}}^{+k_{\rm F}} \frac{ k_{\rm F} \, dk \,e^{ \imath k (y_1-y_2)} }{(\imath(k-\Lambda_{1}) +c)
(\imath(k-\Lambda_{2}) +c)}\nonumber \ , \\
  J^{(0)}_{jj}(\Lambda_{1},\Lambda_{2})&=& \int\limits_{-k_{\rm F}}^{+k_{\rm F}} \frac{ k_{\rm F} \, dk  }{(-\imath(k-\Lambda_{1}) +c)(\imath(k-\Lambda_{2}) +c)}\ , \qquad j=1,2 \nonumber \ ,\\
   J^{(0)}_{12}(\Lambda_{1},\Lambda_{2}) &= & \int\limits_{-k_{\rm F}}^{+k_{\rm F}}\frac{ k_{\rm F} \, dk \,e^{ \imath k (y_1-y_2)-2i\arctan((k-\Lambda_2)/c)} }{(k-\Lambda_{1})^2 +c^2}  \ . \label{eq4.11}
\end{eqnarray}
Moreover $I^{(0)}_{21}(\Lambda_{1},\Lambda_{2})$ $=$ $[I^{(0)}_{12}(\Lambda_{1},\Lambda_{2})]^*$ and
 $J^{(0)}_{21}(\Lambda_{1},\Lambda_{2})$ $=$ $[J^{(0)}_{12}(\Lambda_{2},\Lambda_{1})]^*$. 
The integrals in Eq.~\eqref{eq4.11} reveal that $\tilde{R}_0(y_1, y_2)$ is a function of the
difference $ y_1- y_2$ only, as expected from translation
invariance. For $c\rightarrow 0^+$ the integrals in Eq.~\eqref{eq4.18} can be evaluated using
\begin{eqnarray}
\lim_{c\rightarrow 0^+} \frac{1}{\pi}\frac{c}{(k-\Lambda)^2
+c^2}=\delta(k-\Lambda)\label{eq4.20} \ .
\end{eqnarray} This yields
in the limit of vanishing interaction strength the density-density
correlation function of two free Fermions with momenta
$k_{\uparrow 1}$ and $k_{\uparrow 2}$ is given by Eq.~(\ref{eq4.24}). For hardcore interaction the 
integrals in Eqs.~\eqref{eq4.18} and
\eqref{eq4.11} become trivial and we obtain
\begin{eqnarray}
\lim_{c\rightarrow +\infty}\tilde{R}_0(y_1, y_2) & = &1-\left(\frac{\sin(k_{\rm F}(y_1-y_2))}{k_{\rm F}
(y_1-y_2)}\right)^2.\label{eq4.19}
\end{eqnarray}
This is the density-density correlation function of a non--interacting Fermi-sea with the typical  
decay $\tilde{R}_0 \sim (y_1- y_2)^{-2}$ for large distances.  This large distance behavior remains 
unchanged for finite $c$.

\begin{figure}[htb!]
\begin{center}
\include{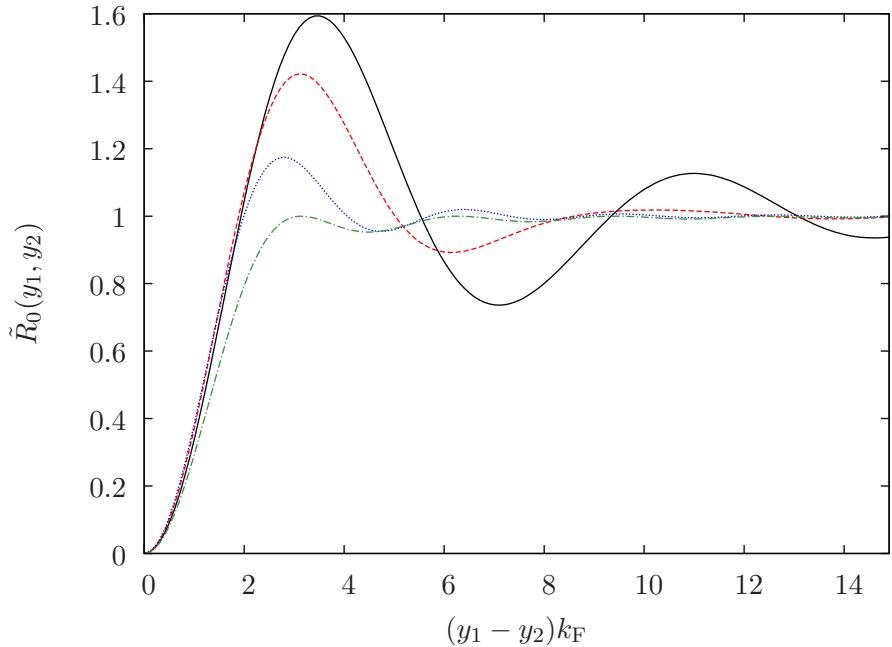}
\caption{Two particle density-density correlation function as function
of the distance $y_1-y_2$ for the quantum numbers
$k_{\uparrow 1}=-k_{\uparrow 2}=\pi k_{\rm F}/8$.
The values for the interaction strength $c$ are: $c=0.05k_{\rm F}$ solid line (black),
$c=0.1 k_{\rm F}$ dashed line (red), $c=0.7 k_{\rm F}$ dotted line (blue)
and $c=+\infty$ dot-dash line (green).}
\label{fig2}
\end{center}
\end{figure}

In Fig.~\ref{fig2} 
$\tilde{R}_0(y_1,y_2; k_{\uparrow 1},k_{\uparrow 2})$ is plotted for the choice  
$k_{\uparrow 1}=-k_{\uparrow 2}=\pi k_{\rm F}/8$ 
and for different values of $c$.
As $c$ varies from zero to infinity the two
particle density-density correlation function undergoes a transition
from that one of two free Fermions to the one of a Fermi-sea. For
other choices of the quantum numbers $k_{\uparrow 1}$ and $k_{\uparrow 2}$ the picture
remains similar.

\subsection{Three particle density-density correlation function}

In contrast to the two particle case now 
$I^{(1)}(\Lambda_{1},\Lambda_{2})$ and 
$J^{(1)}(\Lambda_{1},\Lambda_{2})$ 
represents a $3\times 3$ matrix. The
entries are $I^{(1)}_{ij}$$=$ $I^{(0)}_{ij}$ respectively  
$J^{(1)}_{ij}$$=$ $J^{(0)}_{ij}$  for $i,j=1,2$. Moreover $I^{(1)}_{33}$$=$ $2$ and
\begin{eqnarray} 
J^{(1)}_{33}(\Lambda_{R1},\Lambda_{R2})&=&\left\{ \begin{array}{cc} 
\displaystyle \int\limits_{-k_{\rm F}}^{k_{\rm F}}\frac{dk}{k_{\rm F}} \prod_{n=1}^2e^{ - 2 \imath \arctan((k-\Lambda_n)/c)} \  & {\rm if}\quad  x\in [y_2,y_1]\ ,\cr
			2\ & {\rm otherwise.}\end{array}\right.
\end{eqnarray}
The other entries are given by
\begin{eqnarray} 
\label{eq5.7}
 I^{(1)}_{13}(\Lambda_{1},\Lambda_{2}) & = &\int\limits_{- k_{\rm F}}^{k_{\rm F}}dk k_{\rm F}
\frac{[- \imath(k-\Lambda_{1})+ c \sgn(x-y_1)]
[-\imath(k- \Lambda_{2})+c\sgn(x-y_2)]}{[(k-\Lambda_1)^2
+ c^2][\imath(k-\Lambda_2)+ c]}e^{\imath k(y_1-x)} 
 \ ,\nonumber\\
%%%%%%%%%%%%%%%%%%%%%%%%%%%%%%%%%
 I^{(1)}_{23}(\Lambda_{1},\Lambda_{2}) & = &\int\limits_{-k_{\rm F}}^{k_{\rm F}}dk 
 k_{\rm F}
\frac{[- \imath(k-\Lambda_{1})+ c \sgn( x-y_1)]
[-\imath(k- \Lambda_{2})+ c \sgn( x - y_2)]}{[(k- \Lambda_2)^2
+ c^2][-\imath(k- \Lambda_1)+ c ]}e^{\imath k(y_2- x)} 
\nonumber \ ,\\
%%%%%%%%%%%%%%%%%%%%%%%%%%%%%%%%%%%%%
 J^{(1)}_{13}(\Lambda_{1},\Lambda_{2}) & = &\int\limits_{-k_{\rm F}}^{k_{\rm F}}dk 
 k_{\rm F}
\frac{[- \imath(k-\Lambda_{2})+c\sgn(x-y_1)]
[-\imath(k-\Lambda_{1})+c\sgn(x-y_2)]}{[(k-\Lambda_1)^2
+c^2][\imath(k-\Lambda_2)+c]}e^{\imath k(y_1-x)} 
\nonumber \ ,\\
%%%%%%%%%%%%%%%%%%%%%%%%%%%%%%%%%%%%%
 J^{(1)}_{23}(\Lambda_{1},\Lambda_{2}) & = &\int\limits_{-k_{\rm F}}^{k_{\rm F}} dk
k_{\rm F} \frac{[- \imath(k-\Lambda_{1})+c\sgn(x-y_1)]
[-\imath(k-\Lambda_{2})+c\sgn(x-y_2)]}{[(k-\Lambda_2)^2
+c^2][-\imath(k-\Lambda_1)+c]}e^{\imath k(y_2-x)} 
\nonumber \ ,\\
\end{eqnarray}
moreover $I^{(1)}_{3n}(\Lambda_{1},\Lambda_{2})$ $=$ $[I^{(1)}_{n3}(\Lambda_{1},\Lambda_{2})]^*$,  
$J^{(1)}_{3n}(\Lambda_{1},\Lambda_{2})$ $=$ $[J^{(1)}_{n3}(\Lambda_{2},\Lambda_{1})]^*$ for $n=1,2$. For $c\rightarrow 0^+$ we
make use of relation \eqref{eq4.20} to evaluate the normalization
constant \eqref{eq4.18} and the integrals in Eq.~\eqref{eq5.7}. The outcome is 
\begin{eqnarray}
\lim_{c\rightarrow 0^+}\tilde{R}_1(y_1,y_2,x)&=&
1-\cos\left((k_{\uparrow 1}-k_{\uparrow 2})(y_1-y_2)\right) \label{eq5.8} \ .
\end{eqnarray}
In the hardcore limit  the integrals in
Eqs.~\eqref{eq5.7} become elementary and can be evaluated. 
The resulting expression for
$\tilde{R}_1(\hat{y_1},\hat{y_2},x)$ can be cast into the form
\begin{eqnarray}
\lim_{c\rightarrow+\infty}\tilde{R}_1(y_1,y_2,x)&=&
w(\lambda_1,\lambda_2)
\tilde{R}_{FF}(\hat{y_1},\hat{y_2},x) \  , \label{eq5.26}
\end{eqnarray}
where  we have introduced the quantity
\begin{eqnarray}
 \tilde{R}_{FF}(y_1,y_2,x)&=&
1-\left(\frac{\sin(x-y_1)}{x-y_1}\right)^2
-\left(\frac{\sin(x-y_2)}{x-y_2}\right)^2
-\left(\frac{\sin(y_1-y_2)}{y_1-y_2}\right)^2\\
&&\qquad+2\frac{\sin(x-y_1)}{x-y_1}
\frac{\sin(x-y_2)}{x-y_2}
\frac{\sin(y_1-y_2)}{y_1-y_2} \ \nonumber
\end{eqnarray}
which corresponds to the three particle density-density
correlation function of free Fermions. We recall the definition of 
 $\lambda_n =$ $\lim_{c\to\infty} \Lambda_n/c$ $=$ 
$-\tan( \pi k_{\uparrow\mu}/2k_{\rm F})$. 
It  varies from $-\infty$ to
$+\infty$ as the $k_{\uparrow\mu}$ vary from $k_{\rm F}$ to -$k_{\rm F}$.
The factor $w(\lambda_1,\lambda_2)$ in Eq.~\eqref{eq5.26}  is
given by
\begin{eqnarray}
w(\lambda_1,\lambda_2)= \frac{1-\re ~ uv}{1-\re ~
u}\label{eq5.26-1}
\end{eqnarray}
with
\begin{eqnarray}
u=\frac{(1+\lambda_1^2)(1+\lambda_2^2)}
{(\lambda_1-\lambda_2)^2+4}
\left(\frac{\imath(\lambda_1-\lambda_2)
+2}{(1+\imath\lambda_1)(1-\imath\lambda_2)}\right)^2
\label{eq5.27-1} \ ,\\
v=\frac{(\imath\lambda_1+\sgn
(x-y_2))(\imath\lambda_2+\sgn
(x-y_1))} {(\imath\lambda_1+\sgn
(x-y_1))(\imath\lambda_2+\sgn
(x-y_2))} \ .\label{eq5.27-2}
\end{eqnarray}
According to the equations above $v(\lambda_1,\lambda_2)$ and hence also
$w(\lambda_1,\lambda_2)$ is a piecewise constant function
of $x$. If $x$ lies outside the interval $[y_1,y_2]$,
$v=1$  such that   $
\tilde{R}_{1}(y_1,y_2,x)=
\tilde{R}_{FF}(y_1,y_2,x) $ for
$x\notin(y_1,y_2)$ .
 However, if $x$ lies between the two
spin-up particles the
function $w(\lambda_1,\lambda_2)$ yields a weight for
the density of the Fermi-sea inside the interval
$(y_1,y_2)$ which crucially depends on the quantities
$\lambda_1$ and $\lambda_2$.  We first consider the case 
$k_{\uparrow 1}=-k_{\uparrow 2}$. This implies
$\lambda_1=-\lambda_2$ and Eq.~\eqref{eq5.26-1}
simplifies further
\begin{eqnarray}
\label{eq5.30} 
 w(\lambda_1,-\lambda_1)& = & \frac{(\lambda_1^2 -3)^2}{(1+\lambda_1^2)^2} \ .
\end{eqnarray}
Thus if $k_{\uparrow 1}=\pm k_{\rm F}$ borders the Fermi-sea such that
$\lambda_1=\pm \infty$ we have
$w(\lambda_1,-\lambda_1)=1$ and consequently
 the three particle density-density
correlation function coincides with that one of free Fermions
for all values of
$x$. Most interestingly from Eq.~\eqref{eq5.30} follows that $ w(\lambda_1,-\lambda_1)=0$ for
$\lambda_1=\sqrt{3}$ which corresponds to the choice
$k_{\uparrow 1}=-k_{\uparrow 2}=2k_{\rm F}/3$. This implies that the density
of the Fermi-sea between $y_1$ and $y_2$ vanishes
identically.
%Figure \ref{fig0} shows the plot of
%$w(\hat{\lambda}_1,\hat{\lambda}_2)$ as function of
%$\hat{\lambda}_1$ for different values of $\hat{\lambda}_2$ as
%well as for $\hat{\lambda_2}=-\hat{\lambda_1}$. The picture
%remains similar if both $\hat{J}_1$ and $\hat{J}_2$ are chosen in
%a nonsymmetric way inside the Fermi-sea.

Next we consider the choice
$k_{\uparrow 2}=0$ to be in the core of the Fermi-sea. The quantity
 $w(\lambda_1,0)$ diverges as
$k_{\uparrow 1}\rightarrow \pm k_{\rm F}$ approaches  the border of the
Fermi-sea. Consequently the density of the spin-down particles
in the region
$(y_1,y_2)$ increases as  $k_{\uparrow 1}\rightarrow \pm
k_{\rm F}$ and finally diverges for $k_{\uparrow 1}=\pm k_{\rm F}$.
%\begin{figure}[h!]
%\begin{center}
%\include{3p-density-weigth}
%\caption{Weigth function for the density inside the interval $[\hat{y}_1,\hat{y}_2]$ if $\hat{J}_1=-\hat{J}_2$.}
%\label{fig0}
%\end{center}
%\end{figure}

\begin{figure}
\begin{center}
\include{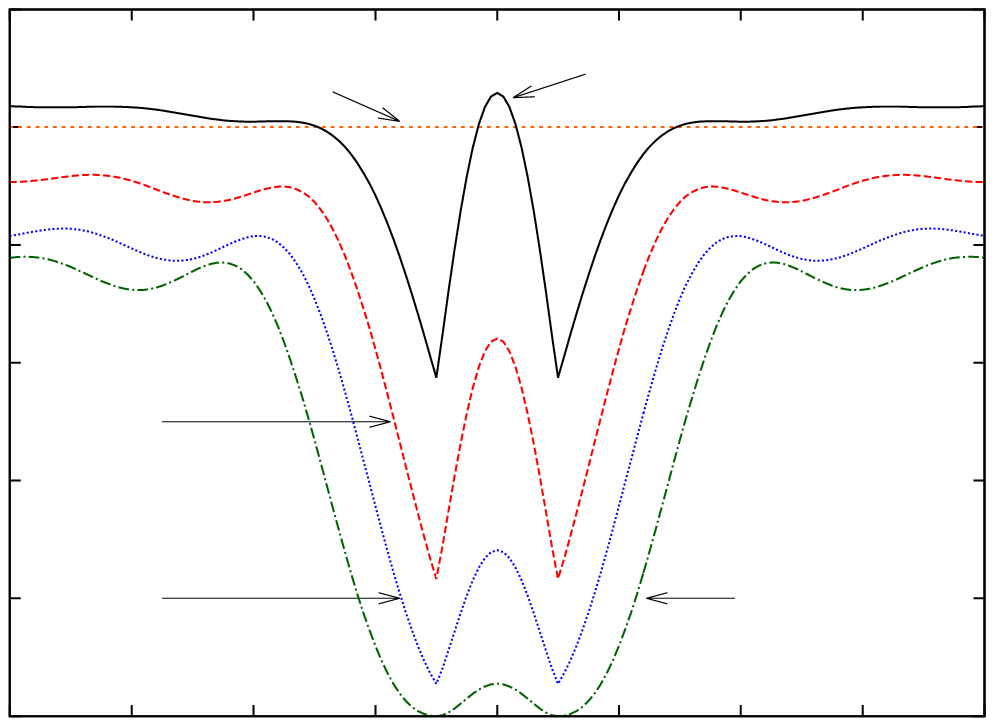}
\caption{ Three particle density-density correlation function for fixed
$y_1=-k^{-1}_{\rm F}$ and $y_2= k^{-1}_{\rm F}$ as function of $x$ for the
quantum numbers $k_{\uparrow 1}=-k_{\uparrow 2}=\pi k_{\rm F}/8$. The values
for the interaction strength are: $c=0$ short dashed line (orange),
$c=0.5 k_{\rm F}$ solid line (black), $c= k_{\rm F}$ dashed line (red),
$c=2 k_{\rm F}$ dotted line (blue) and $c=+\infty$ dot-dash line (green). }
\label{fig3}
\end{center}
\end{figure}

\begin{figure}[h!]
\begin{center}
\include{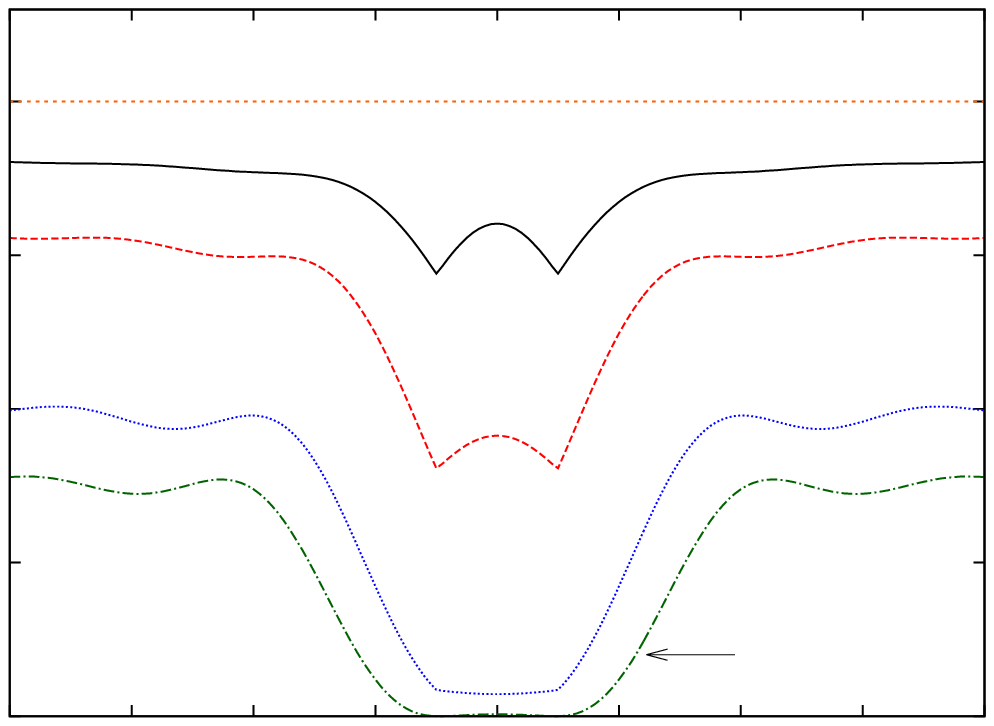}
\caption{ Three particle density-density correlation function for fixed
$y_1= - k_{\rm F}$ and $y_2= k_{\rm F}$ as function of $x$ for the
quantum numbers $k_{\uparrow 1}=-k_{\uparrow 2}=\pi k_{\rm F} /4$. The values
for the interaction strength are: $c=0$ short dashed line (orange),
$c=0.1k_{\rm F}$ solid line (black), $c=0.3 k_{\rm F}$ dashed line (red),
$c=0.5 k_{\rm F}$ dotted line (blue) and $c=+\infty$ dot-dash line (green).
  }
\label{fig4}
\end{center}
\end{figure}

\begin{figure}
\begin{center}
\include{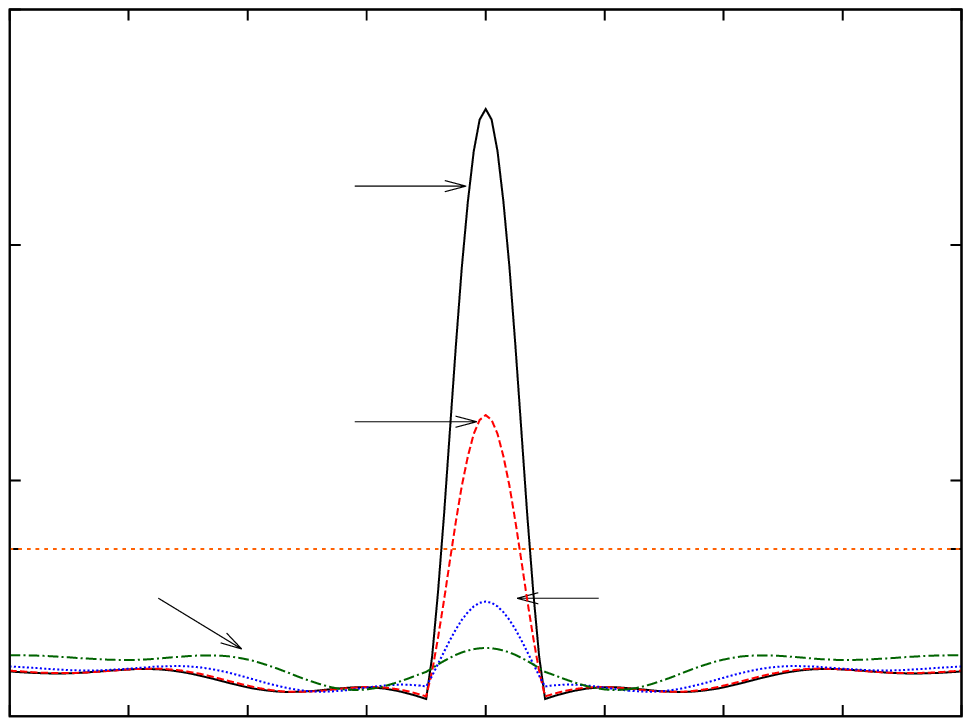}
\caption{ Three particle density-density correlation function for fixed
$y_1k_{\rm F}=-1$ and $y_2k_{\rm F}=+1$ as function of $x$ for the
quantum numbers  $k_{\uparrow 1}= k_{\rm F}$ and $k_{\uparrow 2}=0$ . The values
for the interaction strength are: $c=0$ short dashed line (orange),
$c=5k_{\rm F}$ solid line (black), $c=3k_{\rm F}$ dashed line (red),
$c=k_{\rm F}$ dotted line (blue) and $c=0.1k_{\rm F}$ dot-dash line (green). }
\label{fig5}
\end{center}
\end{figure}

For finite $c$ the integrals \eqref{eq5.7} are evaluated numerically. 
For fixed $y_1$ and
$y_2$ the three particle density-density correlation
function corresponds to the density profile of the Fermi-sea. 
Figure \ref{fig3} shows the plot of $\tilde{R}_1(-k_{\rm F}, k_{\rm F},x)$ for different values
of $c$ where the particles momenta
$k_{\uparrow 1}=-k_{\uparrow 2}=\pi k_{\rm F}/8$ are chosen symmetrically around zero.
While for $c=0$ the density profile of the Fermi-sea is constant it changes when 
interaction is switched on. The density at
the positions of the two spin-up particles decreases as the
interaction increases and finally vanishes for $c\rightarrow
+\infty$.
%For $\hat{c}=0.5$ the density at $\hat{x}=0$ is
%increased with respect to the density at $\hat{c}=0$. For the
%higher values of $\hat{c}$ the density of the Fermi-sea is
%decreased everywhere with respect to the density for $\hat{c}=0$.
Figure~\ref{fig4} shows the same as Fig.~\ref{fig3} but for higher
quantum numbers $k_{\uparrow 1}=-k_{\uparrow 2}=\pi/4$ and slightly
different values of the interaction strength. Comparing
Fig.~\ref{fig4} with Fig.~\ref{fig3} reveals that with increasing
 $c$ the
suppression of the density in 
between the two spin-up particles is
the stronger the higher quantum numbers $k_{\uparrow 1}$ and $k_{\uparrow 2}$
are.
A qualitatively similar picture emerges when the  $k_{\uparrow 1}$ and
$k_{\uparrow 2}$ are inside the Fermi-sea but chosen in a non symmetric
way.
%In Fig.~\ref{fig3} we show the situation where
%$\hat{J_1}=\pi/8$ and $\hat{J}_2=\pi/4$.
%\begin{figure}[h!]
%\begin{center}
%include{3p-density-2d-J=pi-4-pi-8}
%\caption{Three-particle-density-density correlation for fixed $\hat{y}_1=-1$, $\hat{y}_2=+1$ as function of $\hat{x}$ for
%the quantum numbers $\hat{J}_1=\pi/4$ and $\hat{J}_2=\pi/8$.  }
%\label{fig3}
%\end{center}
%\end{figure}
The situation changes rather drastically if one of the quantum numbers,
say $k_{\uparrow 1}$ is set to unity such that it borders the Fermi-sea and the other one
is chosen in the core of the Fermi-sea, that is $k_{\uparrow 2}=0$. Figure \ref{fig5} shows
the  corresponding plots. Now the density between the two spin-up particles is
enhanced with respect to the density in the outer regions. With increasing
interaction strength also the density inside the interval 
$(y_1,y_2)$
increases and finally for $c\rightarrow \infty$ diverges in
accordance with the discussion  following Eq.~\eqref{eq5.30}.

\section{Summary \& Conclusion}\label{sec4}

The Bethe-Ansatz solvable wave function of the Yang Gaudin model was expressed
 as a sum of determinants. Especially in highly imbalanced systems with only a small number of 
minority particles the eigenfunctions acquire a
simple form.

Equilibrium properties of one and two minority Fermions in a sea of majority Fermions were 
investigated. Even for small coupling constants the minority Fermions become strongly entangled 
with the sea particles and form composite particles (polarons). The dispersion relation being 
approximately quadratic in a large range of momenta, estimates for the polaron mass and self 
energy were made.  
Essentially the Fermi sea establishes via $k_{\rm F}^{-1}$ an additional length scale.  On this 
scale the minority Fermions interact among each other giving rise to a non--vanishing interaction 
energy.  

The density-density correlation
function of two minority Fermions shows even for small $c$ the quadratic decay with distance 
which is typical for Fermionic systems. 

Particularly interesting are results obtained for the three--point function. 
They show how the Fermi-sea is locally deformed by the interaction with the minority Fermions 
which act like dynamical impurities. 
The  details of the deformation depend crucially on the free momenta 
$k_{\uparrow 1}$ and $k_{\uparrow 2}$ of the minority Fermions. 
Depending on $k_{\uparrow 1}$ and $k_{\uparrow 2}$ the density 
of sea particles between them can be either suppressed or enhanced.

For infinite strong interaction this becomes most evident.
In this case  our analysis revealed that the density 
in between the two spin-up particles diverges if the quantum
numbers are chosen as $k_{\uparrow 1}=1$ and $k_{\uparrow 2}=0$. 
On the other hand for  $k_{\uparrow 1}=-k_{\uparrow 2}=2k_{\rm F}/3$ the density 
between the two minority Fermions vanishes identically.

\acknowledgments
HK acknowledges financial support from the 
CSIC within the JAE-Doc program  cofunded by the  FSE (Fondo Social Europeo) .  
CR acknowledges support by the "Studienstiftung des deutschen Volkes".
We acknowledge useful discussions with 
F. Guinea, F. Sols and T. Stauber.

\appendix

\section{Proof of Theorem \ref{theo1}}\label{appA}

 In principal one could act directly with the
Hamiltonian on $\Psi({\bf x} ,{\bf k}, {\bf y}, {\bf \Lambda})$ in
Eq.~\eqref{eq1.1} to prove  part one of theorem \ref{theo1}. However, a more handy way to prove
theorem \ref{theo1} is to show that
the wave function fulfills the following conditions:
1.) $ \Psi({\bf x} ,{\bf k}, {\bf y}, {\bf \Lambda})$ fulfills in every sector the free Schr\"odinger equation, 
2.)  $ \Psi({\bf x} ,{\bf k}, {\bf y}, {\bf \Lambda})$ is
    continues everywhere and 3.) let $y_m$ and $x_n$ be adjacent. Then the first derivative of $\Psi({\bf x} ,{\bf k}, {\bf y}, {\bf \Lambda})$ evaluated at $x_n=y_m$ is discontinues such that
\begin{eqnarray}
 \left(\frac{\partial}{\partial x_n} -\frac{\partial}{\partial
    y_m}\right)\Psi({\bf x} ,{\bf k}, {\bf y}, {\bf \Lambda})\Big|_{x_n-y_m=0^-}^{x_n-x_m=0^+}
    &=&4c\Psi({\bf x} ,{\bf k}, {\bf y}, {\bf
    \Lambda})\Big|_{x_n=y_m},\label{A1}
    \end{eqnarray}
where $0^{\pm}$ has the meaning that zero is approached from
above/below. 

As can be seen right from its definition the wave function
\eqref{eq1.1} is continuous and  fulfills  the free Schr\"odinger
within each sector. The continuity of $\Psi({\bf x} ,{\bf k},
{\bf y}, {\bf \Lambda})$  at  $x_n=x_j$ is
obvious, too.
% since there appear no $\sgn$-functions with arguments like $\sgn(x_n-x_j)$.
To show the continuity of the wave function at $x_n=y_m$, it is written 
\begin{eqnarray}
\Psi({\bf x} ,{\bf k}, {\bf y}, {\bf
\Lambda})\Big|_{x_n-y_m=0^{\pm}}\propto \sum_{R\in S_M}
\sgn(R)\prod_{j<l}^{M}\left[ \imath (\Lambda_{Rj} -\Lambda_{Rl})
+2c \sgn(y_l-y_j)\right]\label{A2}\\
\det\left[\ldots\Big|\left[\imath(k_j-\Lambda_{Rm})\pm
c\right]\prod_{s\neq m}^{M}A_j(\Lambda_{Rs},y_m-y_s)e^{\imath
k_jy_m}\Big|\ldots\right]\nonumber \ ,
\end{eqnarray}
where the dots indicate that all other columns of the determinant
\eqref{eq1.2} remain unchanged. Using multilinearity of the determinant 
it is seen that in the difference the $m$-th and the $(M+n)$--th 
column are linearly depended and consequently the
determinant vanishes. This proves the continuity of $\Psi({\bf x}
,{\bf k}, {\bf y}, {\bf \Lambda})$ at $x_n=y_m$. The continuity of the wave-function at 
$y_{\mu}=y_{\nu}$ can be shown similarly.

 To prove the third condition we take the derivative  of
Eq.~\eqref{eq1.0} with respect to $x_n$ on both sides of the border. This yields for the difference 
\begin{eqnarray}
\frac{\partial\Psi({\bf x},{\bf k},{\bf y}, {\bf
\Lambda})}{\partial x_n} \Big|_{x_n-y_m=0^-}^{x_n-y_m=0^+}
\propto\label{A4-1}
\sum_{R\in S_M} \sgn(R)
\prod_{\substack{j<l}}^M\left[\imath(\Lambda_{Rj}-\Lambda_{Rl})
+2c\sgn(y_l-y_j)\right]\\
\det\left[\ldots|2c \imath k_j  \prod_{i\neq m}^M
A_j(\Lambda_{Ri},y_m-y_i)e^{\imath k_j y_m}\Big|\ldots
\right].\nonumber
\end{eqnarray}
Taking the derivative of $\Psi({\bf x},{\bf k},{\bf y}, {\bf
\Lambda})$ with respect to $y_m$ yields
\begin{eqnarray}
&&\left.\frac{\partial\Psi({\bf x},{\bf k},{\bf y}, {\bf \Lambda})}{\partial y_m}
\right|_{x_n-y_m=0^-}^{x_n-y_m=0^+}\propto\label{A5} \\
 &&\qquad \sum_{\substack{R\in  S_M}} \sgn(R)\left[\left(\frac{\partial}{\partial
y_m}\prod_{\substack{j<l}}^M\left[\imath(\Lambda_{Rj}-\Lambda_{Rl})
+2c\sgn(y_l-y_j)\right]\right) \Phi({\bf x},{\bf k},{\bf y,\Lambda})\right. \label{A5-1}\\
&&\qquad \left.\prod_{\substack{j<l}}^M\left[\imath(\Lambda_{Rj}-\Lambda_{Rl})
+2c\sgn(y_l-y_j)\right]\frac{\partial\Phi({\bf
x},{\bf k},{\bf y,\Lambda})}{\partial y_m}\right]_{x_n-y_m=0^-}^{x_n-y_m=0^+}
\label{A5-2}\ .
\end{eqnarray}
Performing the derivative of the prefactor in line \eqref{A5-2}
yield an factor $\delta(y_n-y_m),~n=1,\ldots,M\neq m$
multiplied with $\Phi({\bf x},{\bf k},{\bf y,\Lambda})$ evaluated at
$x_n-y_m=0^{\pm}$. However, as follows from its definition
in Eq.~\eqref{eq1.2} $\Phi({\bf x},{\bf k},{\bf y,\Lambda})$ is
continues at $x_n=y_m$ and hence this term
vanishes when the difference is taken. Thus 
\begin{eqnarray}
\frac{\partial\Psi({\bf x},{\bf k},{\bf y}, {\bf
\Lambda})}{\partial y_m} \Big|_{x_n-y_m=0^-}^{x_n-y_m=0^+}
=\frac{\partial\Phi({\bf x},{\bf k},{\bf y}, {\bf
\Lambda})}{\partial y_m}
\Big|_{x_n-y_m=0^-}^{x_n-y_m=0^+}.\label{A6}
\end{eqnarray}
and it is sufficient to consider the derivative of $\Phi({\bf
x},{\bf k},{\bf y}, {\bf \Lambda})$ with respect to $y_m$.
Using
\begin{eqnarray}
\frac{\partial}{\partial y_m}
\prod_{s=1}^M\left[\imath(k_j-
\Lambda_{R(s)})+2c\sgn(x_l-y_s)\right]e^{\imath
k_j x_l}\Big|_{x_n-y_m=0^\pm}\label{A7}\\
\hspace*{1cm}=\left\{
\begin{array}{ll}
 0 \hspace*{7.5cm}{\rm for} \qquad l\neq n \cr
-2c\delta(x_n-y_m)\prod\limits_{s\neq
m}^MA_j(\Lambda_{Rs},y_m-y_s)e^{\imath k_j y_m}
\hspace*{1cm} {\rm for} \qquad l=n
\end{array}\nonumber
\right.
\end{eqnarray}
as well as properties of the determinant reveals that the
derivative on the first $N$ columns of $\Phi({\bf x},{\bf
k},{\bf y}, {\bf \Lambda})$ with respect to $y_m$ vanishes and
only the derivatives of the last $M$ columns have to taken into
account. We obtain
\begin{eqnarray}
\label{A8} 
&&\frac{\partial\Phi({\bf x},{\bf k},{\bf y}, {\bf \Lambda})}{\partial y_m} \Big|_{x_n-y_m=0^\pm}\propto4c \sum_{l\neq
m}^M\delta(y_m-y_l)\det\left[\ldots\Big|\prod_{\substack{s=1 \cr \neq m,
l}}^MA_j(\Lambda_{Rs},y_l-y_s)e^{\imath k_j y_l}\Big|\ldots\right]+\nonumber\\
 &&\qquad \det\left[\ldots\Big|\imath k_j \prod_{\substack{s=1 \cr \neq m, l}}^MA_j(\Lambda_{Rs},y_m-y_s)e^{\imath k_j
y_m}\Big|\right.
\left.\ldots\Big|\pm c \prod_{\substack{s=1\cr \neq m,
l}}^MA_j(\Lambda_{Rs},y_m-y_s)e^{\imath k_j
y_m}\Big|\ldots\right]\nonumber  .
\end{eqnarray}
Using properties of the determinant it can be shown that the terms
in line \eqref{A8} proportional to $\delta(y_m-y_l)$ vanishes
 when the difference is taken. Hence we have
\begin{eqnarray}
\frac{\partial\Phi({\bf x},{\bf k},{\bf y},
{\bf \Lambda})}{\partial y_m}
\Big|_{x_n-y_m=0^-}^{x_n-y_m=0^+}=\det\left[\ldots\Big|\imath k_j
\prod_{s\neq m, l}^MA_j(\Lambda_{Rs},y_m-y_s)e^{\imath k_j
y_m}\Big|\right.\label{A9}\\
\left.\hspace*{2.5cm}\ldots\Big|2 c \prod_{s\neq m,
l}^MA_j(\Lambda_{Rs},y_m-y_s)e^{\imath k_j
y_m}\Big|\ldots\right]\nonumber \ .
\end{eqnarray}
Observing that
\begin{eqnarray}
\Psi({\bf x},{\bf k},{\bf y}, {\bf \Lambda})
\Big|_{x_n=y_m}\propto\sum_{R\in S(M)} \sgn(R)
\prod_{\substack{j<l}}^M\left[\imath(\Lambda_{Rj}-\Lambda_{Rl})
+2c\sgn(y_l-y_j)\right]\label{A10}\\
\det\left[\ldots\Big| \imath k_j  \prod_{i\neq m}^M
A_j(\Lambda_{Ri},y_m-y_i)e^{\imath k_j y_m}\Big|\ldots
\right]\nonumber
\end{eqnarray}
leads in combination with Eqs.~\eqref{A4-1}, \eqref{A6} and
\eqref{A9} to Eq.~\eqref{A1}. This completes the proof that the
wave function in Eq.~\eqref{eq1.1} is an eigenfunction to the
Hamiltonian \eqref{eq1.0}. The corresponding eigenvalue is given
by $E= \sum_{n=1}^{N+1} k_n^2$, the eigenvalues of the moemntum operator are 
$K= \sum_{n=1}^{N+1} k_n$. This follows from the fact that $\Psi({\bf x},{\bf k},{\bf y}, {\bf \Lambda})$ fulfills the free Schr\"odinger equation in each sector. .

\section{Derivation of two and three particle correlators}\label{appB}

According to
 theorem~\ref{theo1} the eigenfunctions acquire for $M=2$ the form
\begin{eqnarray}
\label{B1}
&&\Psi({\bf x},{\bf k}, y_1,y_2,\Lambda_1,\Lambda_2)\ \propto\ 
\nonumber\\
&&\qquad \sum_{R\in S_2}\sgn(R) \left[\imath (\Lambda_{R1}
-\Lambda_{R2}) + 2c\sgn(y_2-y_1)\right]\Phi({\bf x},{\bf k},
y_1,y_2,\Lambda_{R1},\Lambda_{R2}) \ .
\end{eqnarray}
We use in this appendix the convention
$ A_j(\Lambda)\equiv $ $A_j(\Lambda,1)=$ $ \imath(k_j-\Lambda) +c$. 
To evaluate Eq.~\eqref{eq4.0} for $n=0$ we shift the integration 
variables in Eq.~\eqref{eq4.0} by $x_l \rightarrow x_l +y_1$ and expand the
determinant in Eq.~\eqref{eq1.2} with respect to the last two
columns. This yields
\begin{eqnarray}
\Phi({\bf x}-y_1, y_1,y_2,\Lambda_{R1},\Lambda_{R2})=
\prod_{j=1}^{N+2}e^{\imath k_j y_1}\sum_{n\neq m}^{N+2}
(-1)^{n+m}A_n(\Lambda_{R1})A^*_m(\Lambda_{R1}) e^{\imath k_n
y^-}\label{B5}\\
\hspace*{0cm}\times\det\left[A_j(\Lambda_{R1},x_l
)A_j(\Lambda_{R2},x_l-(y_2-y_1))
e^{\imath
k_jx_l}\right]_{\substack{j=1,\ldots,N+2\neq n,m \cr
l=1,\ldots,N}}\nonumber \ .
\end{eqnarray}
Then by employing properties of the determinant the
integral in the last line of Eq.~\eqref{eq4.0}
 can be cast into the form
\begin{eqnarray}
\int\limits_0^Ldx_1\cdots\int\limits_0^Ldx_N \Phi({\bf x},
y_1,y_2,\Lambda_{R1},\Lambda_{R2})\Phi^*({\bf x},
y_1,y_2,\Lambda_{R^\prime1},\Lambda_{R^\prime2})=N!
\sum_{\substack{n \neq m}}^{N+2} \sum_{\substack{s  \neq t}
}^{N+2}\nonumber
(-1)^{n+m}\\
 (-1)^{s+t}\prod_{j=1}^{N+2}A_j(\Lambda_{R1})A^*(\Lambda_{R^\prime
1}) \frac{A_m^*(\Lambda_{R2}
)A_t(\Lambda_{R^\prime2})}
{A_m(\Lambda_{R1})A^*_t(\Lambda_{R^\prime1})}
e^{\imath(k_n-k_s)y^-}\det\left[Q_{jl}\right]
_{\substack{j,l=1,\ldots,N+2
\cr j\neq n,m ~l\neq s,t}}\label{B6}  ,
\end{eqnarray}
where the quantities $Q_{jl}$ in Eq.~\eqref{B6} are given by
\begin{eqnarray}
Q_{jl}=\int\limits_0^Ldx
A_j(\Lambda_{R2},x-y^-)A_l^*(\Lambda_{R^\prime2},x-y^-)e^{\imath
(k_j-k_l)x}\label{B7} \ .
\end{eqnarray}
The integral is elementary. Evaluating it we obtain
\begin{eqnarray}
Q_{jl}&=& \left[LA_j(\Lambda_{R2})A_l^*(\Lambda_{R^\prime2}) +2c
\imath \left(\Lambda_{R2}-\Lambda_{R^\prime2}\right)y^-\right] \delta_{jl}\label{B8}\\
&&\qquad -2c\left(1-\frac{\Lambda_{R2}
-\Lambda_{R^\prime2}}{k_j-k_l}\right)e^{\imath(k_j-k_l)y^-}(1-\delta_{jl})
+{\rm B.T.}\nonumber \ ,
\end{eqnarray}
where ${\rm B.T.}$ stands for the expression
\begin{eqnarray}
{\rm B.T.}&=&\frac{1}{\imath(k_j-k_l)}
\left[\left(\left(k_j-\Lambda_{R2}\right)\left(k_l-
\Lambda_{R^\prime2}\right)
+c^2\right)\left(e^{\imath(k_j-k_l)L}-1\right)\right.\label{B9}\\
 &+&\left.\imath
c\left(k_j -k_l -\Lambda_{R2}+\Lambda_{R^\prime2}\right)
\left(e^{\imath(k_j-k_l)L}+1\right)\right](1-\delta_{jl})
\nonumber
\end{eqnarray}
which corresponds to terms which arise from the boundaries when
 the off-diagonal terms with $j\neq l$ in Eq.~\eqref{B7} are integrated. However, using the
Bethe-Ansatz equations it shown these terms vanish identically as
a consequence of translational invariance.
The expression~\eqref{B8} reveals that the diagonal terms where
$j=l$ scale like $L$ while the off-diagonal terms scale like $c$.
Therefore, in the thermodynamic limit the off-diagonal terms are
negligible and the entries can be approximated by the diagonal
terms only. In leading order of $L$ therefore 
\begin{eqnarray}
 \hspace*{0cm}\det\left[Q_{jl}\right]_{\substack{j=1,\ldots,N+2\neq n,m\cr
l=1,\ldots,N+2\neq s,t}}=\left(\prod_{j\neq n,m}^{N+2}L
A_j(\Lambda_{R2})A_j^*(\Lambda_{R^\prime2})
\right)\det
\left[\begin{array}{cc}
\delta_{n s} & \delta_{n t} \cr
\delta_{m s} & \delta_{mt }
\end{array}\right] \label{B10} .
\end{eqnarray}
Combining the expression \eqref{B10} with Eq.~\eqref{B6}
the two particle density-density correlation function acquires the
form
\begin{eqnarray}
 R_0(y_1,y_2)&\propto &\sum_{R,R^\prime \in
S_2}\sgn(R+R^\prime)\left[\imath (\Lambda_{R1} -\Lambda_{R2}) +
2c\right]\left[-\imath ( \Lambda_{R^\prime1}
-\Lambda_{R^\prime2}) + 2c\right]\label{B11}\\
&& \sum_{n\neq m}^{N+2}\sum_{s\neq
t}^{N+2}\frac{A^*_m(\Lambda_{R2})A_t(\Lambda_{R^\prime
2})e^{\imath(k_n- k_s)y^-}}
{A_m(\Lambda_{R1})A^*_t(\Lambda_{R^\prime1})A_n(\Lambda_{R2})
A_m(\Lambda_{R2})
A^*_n(\Lambda_{R^\prime 2})A^*_m(\Lambda_{R^\prime2})}
\det
\left[\begin{array}{cc}
\delta_{n s} & \delta_{n t} \cr
\delta_{m s} & \delta_{mt }
\end{array}\right] 
\ .\nonumber
\end{eqnarray}
Since the terms in the second line of Eq.~\eqref{B11} factorize
in all summation indices we can write each term into the
corresponding row or column of the determinant. Furthermore, due
to the Kronecker-$\delta$'s two of the four summations drop out.
Thus the second line of the equation above can be expressed by a
determinant whose entries are one-fold sums. Now it is straightforward 
to take the thermodynamic limit. Assuming the Fermi-sea to
be in the ground state i.e. at zero temperature, the quasi-momenta
distribute themselves uniformly between $\pm k_{\rm F}$ with a
density $\varrho(k)=L/(2\pi)$. In the usual way replace the sums
in Eq.~\eqref{B11} over the quasi-momenta by integrals. Then
Eq.~\eqref{B11} can be cast into the form
\begin{eqnarray}
 R_0(y_1,y_2)&\propto& \left[(\Lambda_1-\Lambda_2)^2+4c^2\right] \det\left[I^{(0)}(\Lambda_{1},\Lambda_{2})\right] \nonumber\\
      &&\qquad- 2\re \left(\imath (\Lambda_1-\Lambda_2)+2c\right)^2 \det\left[J^{(0)}(\Lambda_{1},\Lambda_{2})\right] \ .
\end{eqnarray}
The two 
matrices $I^{(0)}(\Lambda_{1},\Lambda_{2})$ $=[I^{(0)}_{jl}(\Lambda_{1},
\Lambda_{2})]_{j,l=1,2}$  and $J^{(0)}(\Lambda_{1},\Lambda_{2})$ $=$$[J^{(0)}_{jl}(\Lambda_{1},
\Lambda_{2})]_{j,l=1,2}$ are given in
\eqref{eq4.11}. 
Together with the normalization this yields to the form
\eqref{eq4.3-1} of the two particle density-density correlation
 function. It remains to determine the normalization constant
$\mathcal{R}$. The normalization condition reads
\begin{eqnarray}
 4=\int\limits_0^Ldy_1\int\limits_0^Ldy_2 \ R_0(y_1,y_2) \ .
\label{B15}
\end{eqnarray}
To evaluate it we use the form \eqref{B11} of $R_0(y_1,y_2)$. The
integration over the exponential there yields in leading order
of $L$
\begin{eqnarray}
 \int\limits_0^Ldy_1\int\limits_0^Ldy_2 e^{\imath(k_n-k_s)(y_2-y_1)}
=L^2\delta_{ns} \label{B16}
\end{eqnarray}
and thus only the diagonal terms of the entries contribute to
the normalization. This immediately leads to
Eq.~\eqref{eq4.18}.

\bibliography{two-particle2}

\end{document}